\documentclass[%
 reprint,
superscriptaddress,
 amsmath,amssymb,
 aps,
floatfix,
]{revtex4-1}
\usepackage{enumerate}
\usepackage{ulem}
\usepackage{float}
\usepackage{xcolor}
\usepackage{graphicx}
\usepackage{dcolumn}
\usepackage{bm}

\usepackage{wrapfig}

\begin{document}

\preprint{APS/123-QED}

\title{Gauge mean-field theories of the underscreened Kondo lattice}

\author{Ewan Scott}
\affiliation{Department of Mathematics, University College London, Gordon St., London WC1H 0AY, United Kingdom}

\author{Michal Kwasigroch}
\affiliation{Department of Mathematics, University College London, Gordon St., London WC1H 0AY, United Kingdom}
\affiliation{Trinity College, Cambridge, CB2 1TQ, United Kingdom}
\date{\today}

\begin{abstract} Various mean-field  decoupling schemes have been introduced thus far to study the important and challenging problem of the spin-$1$ underscreened Kondo lattice where magnetic order can coexist with Kondo hybridization. We use a single control parameter $N$ and an unbiased variational ansatz to unify and connect the previously proposed decouplings to standard Read-Newns theory, where fluctuations are small in $1/N$.
We compute the corrections around the large-$N$ limit. In particular, we make contact with  Nozières strong-coupling theory by finding the residual ferromagnetic Hund interaction that decays logarithmically in the case of a heavy-fermion metal. We map out the ground state phase diagram as a function of $N$ and the Kondo coupling. We find crucial differences between the previously proposed mean-field theories in the strength of the hybridization and total magnetization of the coexistent phase.  We show that, within our unifying variational theory, the previously proposed decouplings correspond to either taking $N=2$ from the start, or performing a $1/N$ expansion and then extrapolating to $N=2$. Finally, we summarize the generalization of our theory to $S>1$.
\end{abstract}

\maketitle

\section{Introduction}
The coexistence of Kondo and magnetic fluctuations is responsible for a rich variety of unusual phenomena, especially in the vicinity of the quantum critical point that separates the heavy Fermi liquid and magnetically ordered phases. These include heavy-fermion superconductivity in actinide~\cite{Ce_super} or lanthanide~\cite{URhGe,UTe2} compounds, or strange metal behaviour in CeRh$_6$Ge$_4$~\cite{Strange_Metal}. If, additionally, an external magnetic field is applied, Kondo hybridisation and magnetic interactions can give rise to unconventional high-field re-entrant superconductivity, or metamagnetic transitions~\cite{URhGe_metamagnetic}.

\begin{figure}
    \centering
    \includegraphics[width=0.48\textwidth]{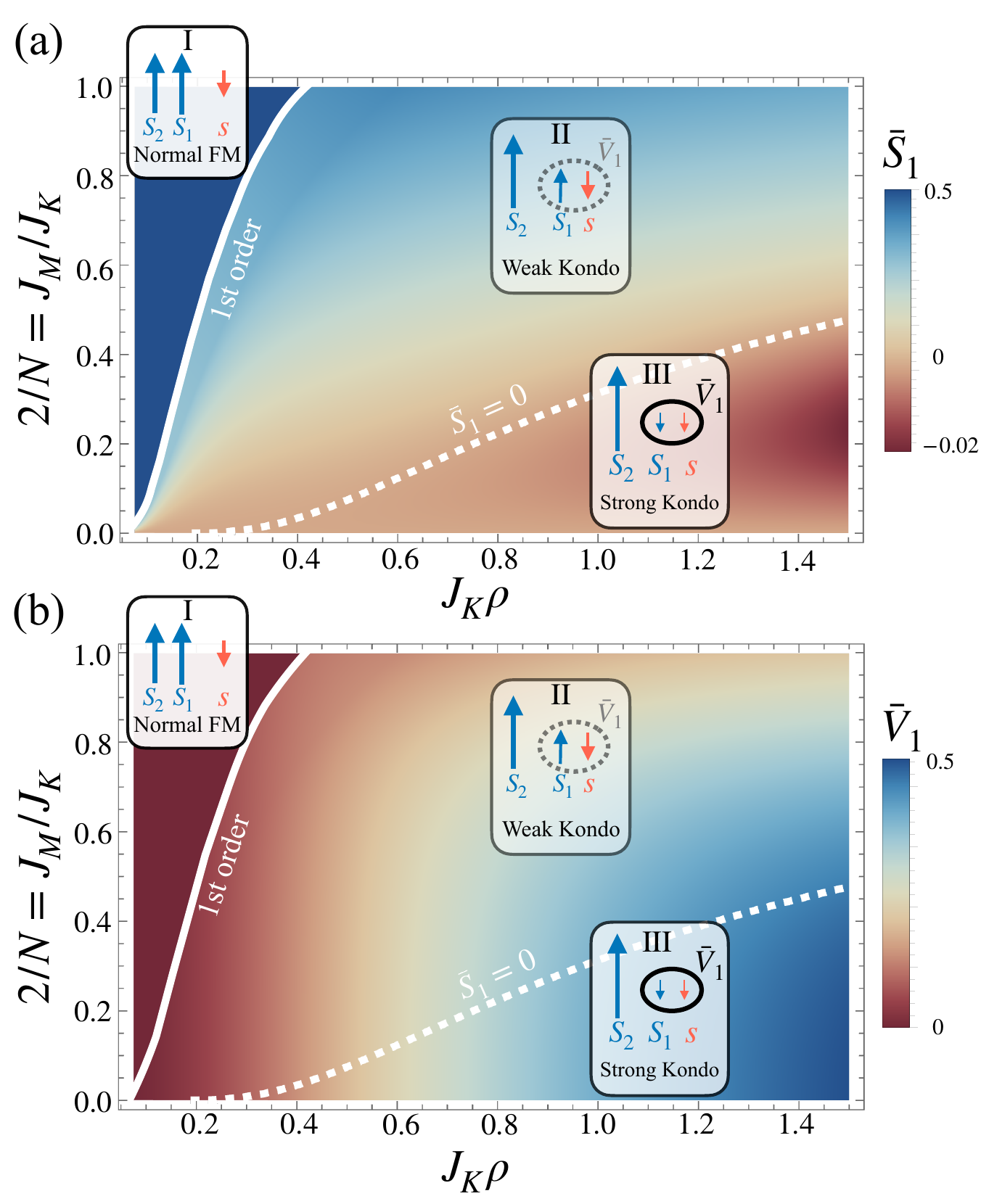}
    \caption{Ground state phase diagrams showing three phases: I -- The normal FM state with local moments maximally polarized $\bar{S}_{1,2}=1/2$ and no hybridization $\bar{V}_1=0$, II -- Weak Kondo ferromagnet with magnetization of the hybridized $f$-moment $\bar{S}_1>0$ and weak hybridization $\bar{V}_1>0$, III -- Strong Kondo ferromagnet with weak magnetization of the hybridized $f$-moment $\bar{S}_1<0$ and strong hybridization, $\bar{V}_1>0$. The unhybridized $f$-moment has a magnetization $\bar{S}_1=1/2$ in all phases. Plot (a) shows the magnetic moment of the hybridized $f$-electron $\bar{S}_1$ as a function of the dimensionless Kondo coupling $J_K\rho$ and the relative strength of the coupling in the magnetic exchange channel $J_M=\frac{2J_K}{N}$. (b) plots the hybridisation, $\bar{V}_1$, as a function of the same. The transition from non-hybridised magnetism to weak Kondo is first order. The transition from weak to strong Kondo is a crossover with the hybridisation and magnetisation changing continuously.}
    \label{fig:phasedia}
\end{figure}

Interest in studying the coexistence is borne out of the simultaneous experimental observation of magnetic order and Kondo signatures  in a wide array of heavy fermion metals, e.g., the order is accompanied by a pronounced coherence maximum in the temperature-dependence of the resistivity, or the ordered moment sizes are significantly reduced below what is expected for the ground state doublet \cite{Brando,7zsl-4497}. 

In spite of the rich phenomenology and direct experimental observation~\cite{UAs2}, the theoretical description of coexisting Kondo $c-f$ hybridisation and RKKY-induced order of local moments, remains one of the major outstanding challenges of heavy-fermion physics. While the Read-Newns large-$N$ theory has given a lot of insight in describing the hybridisation of localised electrons with the conduction band and formation of heavy-fermion bands \cite{PhysRevLett.57.877,NRead_1983}, 
magnetic order is absent in the standard application of the large-$N$ theory to a spin$-1/2$ Kondo lattice in the $N\rightarrow\infty$ limit. Exact methods such as NRG, DMRG, and Bethe ansatz calculations are either restricted to a small number of impurities or the special case of 1D, where the Doniach diagram is inverted: there is no magnetic order in the RKKY dominated phase at weak Kondo coupling and the strongly-coupled phase is actually ferromagnetic \cite{McCulloch2002,Masui2022}. With empirical or mean-field type methods, progress has been made by describing the system as a composite of two fluids, with one component corresponding to unhybridised local moments that can order, and the other to the coherent heavy-fermion Fermi liquid. This has been done phenomenologically~\cite{PhysRevLett.92.016401} and microscopically using supersymmetric spin representations~\cite{PhysRevB.93.035120,PhysRevB.62.3852}. Finally, Monte Carlo calculations have identified a large region of coexistence of magnetic order and $c-f$ hybridisation \cite{PhysRevResearch.2.013276}. However, as the calculations are restricted to the insulating limit, the heavy-fermion Fermi liquid phase cannot be accessed.

The underscreened Kondo lattice (UKL) \cite{Perkins_2007, PhysRevB.87.205107}, in which the number of conduction channels available to screen the impurity moment is restricted, offers a natural setting where the coexistence of magnetic order and Kondo hybridization can be investigated, particularly in the interesting strong-coupling limit. This is often the case for heavy fermion magnets, where the coherence temperature can be an order of magnitude larger than the ordering temperature \cite{7zsl-4497}. 
As a result of the restriction, a part of the moment is left unscreened and able to magnetise through the RKKY interaction, even when the Kondo coupling is strong. The spin-1 UKL model has had considerable success in describing uranium based compounds \cite{Perkins_2007, PhysRevB.76.125101}, where two Hund-coupled 5$f$ electrons can form an effective $S=1$ moment and only one conduction channel has sufficiently strong coupling to hybridize. Beyond the heavy-fermion context, the UKL has been used to shed light on the controversy of perovskite nickelates \cite{PhysRevB.87.205107}.

\par

The UKL is a challenging problem that evades exact treatment beyond 1D. The problem has been addressed with various approximate methods such as static mean-field theory \cite{7zsl-4497, PhysRevB.87.205107,Perkins_2007, PhysRevB.76.125101,Thomas2011, THOMAS2014}, or dynamical mean-field theory (DMFT) \cite{Golez2013}. Because of the presence of both Kondo hybridization and RKKY magnetic channels, which involve competing mean-field decouplings, the optimal decoupling that captures the coexistent phase is not clear. For this reason, various decouplings have been proposed in the past \cite{7zsl-4497, PhysRevB.87.205107,Perkins_2007, PhysRevB.76.125101}. Here, we use a control parameter $N$ and a single variational ansatz, introduced in our earlier work \cite{7zsl-4497}, to unify and connect most of these approaches to the standard Read-Newns theory for the spin-1/2 Kondo lattice. We compute fluctuation corrections to first order in $1/N$. We find that the two main mean-field decoupling schemes in use correspond to: (i) setting the control parameter $N=2$ at the outset; (ii) performing a $1/N$ expansion to first order and then extrapolating to $N=2$. We show that the resulting phase of coexisting magnetic order and Kondo fluctuations possesses a different internal structure in the two cases, such as the relative impurity and conduction electron magnetizations.

Large-$N$ Read-Newns theory is an example of gauge mean-field theory (gMFT). gMFT has successfully characterised many strongly correlated ground states that do not possess a physical (gauge invariant) order parameter, e.g., heavy-fermions \cite{NRead_1985,PhysRevB.28.5255,PhysRevB.101.075133,Coleman_2015}, quantum spin liquids \cite{PhysRevLett.118.087203, PhysRevB.87.205107}.  It achieves this by approximating the action with its value at the $N\rightarrow\infty$ saddle-point that is characterised by a fixed gauge ($N$ is now the general gMFT control parameter). Gauge-symmetry-restoring fluctuations are quenched, and consequently, the constraint keeping the ground state to the physical gauge sector is relaxed. Coherence between different gauge sectors is established and the groundstate is well approximated by a product state. Although the gMFT groundstate is a product state, it actually represents a strongly-entangled state in the physical gauge sector. (Non-physical gauge sectors can be projected out with a Gutzwiller operator.) The fractional violation of the constraint scales as $1/\sqrt{N}$, which is why gMFT becomes exact as  $N\rightarrow\infty$. Importantly, gMFT correctly describes many groundstate properties even in cases where the general control parameter $N$ is not strictly large, often with quantitative accuracy  \cite{PhysRevB.95.134439, PColeman_1986}. The system can be thought of as flowing in the RG sense towards the $N\rightarrow\infty$ fixed point \cite{PhysRevB.95.134439}. It is important to remember that gMFT has been far less successful at non-zero temperatures due to rapidly growing thermal fluctuations of the gauge fields that restore the gauge symmetry \cite{PhysRevB.28.5255,Coleman_2015}.

We introduce the spin-1 UKL model in Sec. \ref{sec:model} before discussing the fixed-gauge $N\rightarrow \infty$ saddle point in Sec. \ref{sec:saddle-point}. We then introduce the unbiased variational ansatz that connects various mean-field decouplings in Sec. \ref{sec:variational}. Finally, we discuss the resulting ground state phases in Sec. \ref{sec:ground state}.

\section{Model} \label{sec:model}The underscreened $S=1$ Kondo lattice is described by the following Hamiltonian
\begin{equation}\label{Hamil} 
H=J_K\sum_{i}\mathbf{S}(\mathbf{r}_i)\cdot\mathbf{s}(\mathbf{r}_i)+\sum_{\mathbf{k}\sigma}\epsilon_\mathbf{k} c^\dagger_{\sigma}(\mathbf{k})c_{\sigma}(\mathbf{k}),
\end{equation}
where $J_K>0$ is the Kondo coupling between impurity spin $\mathbf{S}_i$, and the conduction electron spin $\mathbf{s}_i$. We are approximating the conduction band with a sharp cutoff $|\epsilon_\mathbf{k}|\leq\Lambda$ and a constant density of states $\rho=\frac{1}{2\Lambda}$. 

Although we shall focus on the $S=1$ UKL, our theory can be straightforwardly generalised to $S>1$ (see App. \ref{app:generalisation} for futher details). We write the $S=1$ impurity moment in the Schwinger fermion representation \cite{PhysRevB.76.125101, PhysRevB.87.205107},
\begin{equation}\label{abrikosov}
S^{\eta}(\mathbf{r}_i)= \sum_{a=1,2} S^{\eta}_{a}(\mathbf{r}_i) =
\frac{1}{2}\sum_{a\sigma\sigma'} f_{a\sigma}^\dagger(\mathbf{r}_i) \sigma^{\eta}_{\sigma\sigma'} f_{a\sigma'}(\mathbf{r}_i),
\end{equation}
where $a=1,2$ indexes the two spin-1/2 fermions, $\sigma^{x,y,z}$ are Pauli matrices and 
$f_{a\sigma}(\mathbf{r}_i)$ obey the usual fermionic anticommutation relations. In this representation, the Hamiltonian can be written (up to a constant) as 
\begin{align} \label{FermiHam}
H_N=&\frac{J_K}{N}\sum_{a\alpha\sigma\beta\sigma' i}f^\dagger_{a\alpha\sigma }(\mathbf{r}_i)f_{a\beta\sigma' }(\mathbf{r}_i) c^\dagger_{\beta \sigma' }(\mathbf{r}_i)c_{\alpha\sigma }(\mathbf{r}_i)
\nonumber\\
&+\sum_{\mathbf{k}\alpha}\epsilon_\mathbf{k} c^\dagger_{\alpha\sigma}(\mathbf{k})c_{\alpha\sigma}(\mathbf{k}),
\end{align}
where, as in our earlier work \cite{7zsl-4497},
 we have promoted the total number of fermionic spin flavours to $N$ by introducing $N/2$ replicas of up spins ($\sigma=1/2$) as well as $N/2$ replicas of down spins ($\sigma=-1/2$), indexed by $\alpha$ or $\beta$. This generalises both $f_1$ and $f_2$ $SU(2)$  spins to $SU(N)$ spins. $N$ will be the gMFT control parameter. The following constraints isolate the physical spin degrees of freedom
\begin{eqnarray}\label{constraint}
    \hat{n}^f_i&:=& \sum_{a} \hat{n}^f_{a i} = \sum_{a\alpha\sigma} f^{\dagger}_{a\alpha\sigma}(\mathbf{r}_i) f_{a\alpha\sigma}(\mathbf{r}_i) = N,
    \nonumber\\
   \hat{{T}}^{\eta}_i &:=& \frac{1}{2}\sum_{a a' \alpha\sigma} f^{\dagger}_{a\alpha\sigma}(\mathbf{r}_i) \tau^{\eta}_{a a'} f_{a'\alpha\sigma}(\mathbf{r}_i) = 0,
\end{eqnarray}
where $\tau^{x,y,z}$ are Pauli matrices that act on the 2-dimensional $a$-subspace (i.e. the fermion 'flavour' $a=1,2$ subspace, rather than the fermion spin $\sigma=\pm\frac{1}{2}$ subspace). The constraints $\hat{n}^f_i$ and $\hat{T}^z_i$ guarantee that there are always $\hat{n}^f_{1 i}=\frac{N}{2}$ $f_{1}$-fermions and  $\hat{n}^f_{2 i}=\frac{N}{2}$ $f_{2}$-fermions on each lattice site. The constraints $\hat{T}^{x,y}$, on the other hand, represent an infinite Hund's coupling between the $SU(N)$ $f_1$ and $f_2$ spins, which for $N=2$ projects out the spin singlet state.  As the constraints commute with $H_N$, the energy eigenstates are invariant (in the sense that the eigenvalues remain the same) under the direct product of the following local $U(1)$ and $SU(2)$ symmetry transformations
\begin{eqnarray}
    \hat{U}(\theta_i) &=&\exp\left[i\left(\hat{n}^f_i-N\right)\theta_i\right],
\nonumber\\
 \hat{U}(\phi^{\eta}_i) &=&
    \exp\left[i \phi^{\eta}_i \hat{T}^{\eta}_i \right].
\end{eqnarray}
These make up the local $U(2)\sim U(1)\times SU(2)$ gauge group. Fluctuations of the $U(1)$ and $SU(2)$ gauge fields $\theta_i$, $\bm{\phi}_i$ respectively, enforce the constraints in Eq.~\ref{constraint}. It should also be noted that the following Gutzwiller projection operator can be used to project out states that violate the constraints, e.g., to heal the MF state by projecting onto the physical gauge sector
\begin{align}
   \prod_i\int_0^{2\pi} \frac{{\rm d}\theta_i}{2\pi}
   \int_{|\boldsymbol{\phi_i}| <2\pi}
   \frac{{\rm} d^3 \boldsymbol{\phi_i} }{8\pi^2} &\frac{1-\cos|\boldsymbol{\phi_i}|}{|\boldsymbol{\phi_i}|^2}\times
   \nonumber\\
   &\exp\left(
   i\theta_i\left(\hat{n}^f_i-N
   \right)+
   i\boldsymbol{\phi}_i \cdot \hat{\boldsymbol{T}_i}
   \right), \label{eq:Gutzwiller}
\end{align}
where $\boldsymbol{\phi}_i=(\phi_i^x,\phi_i^y,\phi_i^z)$.

\section{Fixed-gauge $N\rightarrow\infty$ saddle point} \label{sec:saddle-point} Read-Newns large-$N$ theory introduces the hybridisation field
 \begin{equation}
     \begin{pmatrix}
         \hat{V}_{1i}\\ \hat{V}_{2i}
     \end{pmatrix}=\frac{1}{N}\sum_{\alpha\sigma} 
     \begin{pmatrix}
          f^\dagger_{1\alpha\sigma}(\mathbf{r}_i)c_{\alpha\sigma}(\mathbf{r}_i)\\
         f^\dagger_{2\alpha\sigma}(\mathbf{r}_i)c_{\alpha\sigma}(\mathbf{r}_i)
     \end{pmatrix},
 \end{equation}
as the order-parameter with local $U(2)$ gauge symmetry. This is a gauge-dependent operator and fluctuations of the gauge fields reduce its expectation value to zero -- a consequence of Elitzur's theorem which prohibits the breaking of local gauge symmetry.
Nonetheless, we can exploit the $U(2)$ symmetry to perform a change of variables that achieves $\langle \hat{V}_{1i}\rangle>0$ and $\langle \hat{V}_{2i}\rangle=0$ (in terms of the new variables). This procedure is known as gauge-fixing and leaves physical operators unchanged. Details can be found in App. \ref{app:gauge}.  The $f_{2}$-moments thus remain unhybridised in the fixed gauge, which also reflects the underscreened nature of the model. $U(2)$ symmetry has been reduced to a $U(1)$ symmetry corresponding to the free fluctuations of the phase of the unhybridised $f_2$ fermions.

One should explore the fluctuations around the fixed-gauge saddle point to check the validity of the above approach. The $\lambda_{1i}:=i(\theta_i +\phi^z_i)$ gauge field develops a non-zero stiffness, relaxing the constraint on $\hat{n}^f_{1i}$, that can be viewed as an infinitely strong Hubbard-like interaction between $f_1$-fermions, instead giving rise to {\it finite} Hubbard-like Fermi-liquid interactions. Similarly, the $\phi^{x,y}_i$ gauge fields also acquire a non-zero stiffness, which  reduces the infinitely strong Hund exchange between $f_1$- and $f_2$- spins to a {\it finite} residual Hund-like ferromagnetic interaction that, away from the insulating limit, scales logarithmically to zero with temperature. The fluctuation-generated corrections to the mean-field Hamiltonian include
\begin{align}\label{corrections}
 &\Delta H_{\rm fluc.} = \frac{1}{2}G_{\lambda\lambda}(0) \sum_i \left(n^f_{1i}-\frac{N}{2}\right)^2 +
   \nonumber\\
   &G_{\phi^{\dagger}\phi}(0)\sum_{i,\alpha\sigma\beta\sigma'} f_{1\alpha\sigma}^\dagger(\mathbf{r}_i)  f_{1\beta\sigma'}(\mathbf{r}_i) f_{2\beta\sigma'}^\dagger(\mathbf{r}_i) f_{2\alpha\sigma}(\mathbf{r}_i),
\end{align}
where the gauge-field Green's functions in the low-energy limit scale as $G_{\lambda\lambda}(0)\sim T_K/N$ and  $G_{\phi^{\dagger}\phi}(0)\sim T_K(N \log(T/T_K))^{-1}<0$, and a detailed calculation is included in App. \ref{app:gauge}. The latter Green's function has the same form as the residual ferromagnetic exchange between a single underscreened impurity and conduction electrons \cite{PhysRevB.68.220405}, but we believe this is the first time the residual ferromagnetic exchange is derived between a lattice of underscreened moments and the heavy-fermion Fermi liquid (the $f_1$-moments in the above equation are now part of the heavy-fermion band via $c-f$ hybridisation).  

The constraints are thus violated in the gMFT approximation, but importantly fractional fluctuations in the constraints $\frac{1}{N}\langle  (\hat{T}^{\eta})^2 \rangle^{1/2}\sim 1/\sqrt{N}$  become small as $N\rightarrow \infty$, when gMFT becomes exact.
We also point out that since gMFT captures the behaviour of the spin-1/2 Kondo-lattice, even though the residual Hubbard-like interactions in Eq.~\ref{corrections} are not technically small for $N=2$, it should also capture the behaviour of the underscreened spin-1 Kondo lattice, as the residual Hund-like exchange scales logarithmically to zero with temperature for all $N$, i.e., as $T\rightarrow 0$, the $\phi_i^{x,y}$ gauge fields become infinitely stiff.

The success of gMFT in describing the Kondo-lattice is linked to the orthogonality catastrophe and the slow power-law decay of coherence of the hybridisation field with time.  As it is the fluctuations in the phase of $\hat{f}_{1\alpha i}$ (that correspond to dynamic fluctuations of $\lambda_{1i}$), as well as its azimuth in the $f_1$-$f_2$ plane (that correspond to dynamic fluctuations of $\phi_i=\phi_i^x-i\phi_i^y$), that lead to the decay of $\langle\hat{V}_{1i}\rangle$ in time,  the decoherence of the hybridisation field is captured by the correlator $\langle f_{1 \alpha i}^{\dagger}(t) f_{1 \alpha i}(0)\rangle \propto t^{-1/N} $, which is directly related to the post-ionisation decay of the overlap with the initial state in the X-ray orthogonality catastrophe \cite{PhysRevLett.18.1049,PhysRev.178.1097}. The exponent of $1/N$ for the underscreened spin-1 Kondo impurity is the same as that for the spin-1/2 Kondo impurity. This is because the $\phi_i$ gauge field becomes infinitely stiff in the ground state and does not add to the decoherence of the hybridisation field, which is thus entirely caused by the fluctuations of the $\lambda_{1i}$ gauge field. 
 
\section{Unbiased variational approach} \label{sec:variational}
If the unscreened $f_2$-moments only couple to conduction electrons through the Kondo channel, i.e. via the hybridisation field $\hat{V}_{2i}$, as would be the case for a single $S=1$ Kondo impurity, they completely decouple from the system in the ground state (the residual ferromagnetic interaction scales logarithmically to zero as the temperature is lowered). However, the situation will be drastically different if coupling through the magnetic RKKY channel cannot be neglected, as would be the case in the presence of long-range magnetic order. While $U(2)$ gauge symmetry can be used to decouple $f_2$-moments from the hybridisation field entirely, and hence from conduction electrons in the Kondo channel, the magnetic mean-field order parameter is a physical gauge-invariant operator, invariant under $U(2)$ transformations, and hence $f_2$-moments cannot be decoupled from conduction electrons in the RKKY channel. In other words, if magnetic order is present, RKKY interactions between $f_2$-moments, generated by the conduction electrons in the RKKY channel need to be included. This was done with success in Ref. \cite{PhysRevB.87.205107}, where an effective Hamiltonian was derived in the degenerate subspace of $f_2$-moments by treating the difference between $H_N$ and the gMFT Hamiltonian with $\langle\hat{V}_{2i}\rangle=0$ as a perturbation (this is essentially the exchange term between $f_2$-moments and conduction electrons).  The phases obtained after the degeneracy is lifted agreed well with exact DMRG results. 

We can also obtain the RKKY interactions as follows. The logarithmic decoupling of the $f_2$-moments in the Kondo channel arises because of the quenching of slow azimuthal fluctuations of the hybridization order parameter away from $\hat{V}_{1i}$ and in the direction of $\hat{V}_{2i}$. (The decoherence of $\langle\hat{V}_{1i}\rangle$ in time is caused entirely by the fluctuations of the phase of $\hat{V}_{1i}$ that correspond to dynamic fluctuations of $\lambda_{1i}$.) Incoherent, fast fluctuations of  $\hat{V}_{2i}$, on the other hand, are still allowed and generate an exchange interaction between the $f_2$-moments and conduction electrons that scales like $1/N$ and can be treated perturbatively. Treating the $f_2$-moments in the static (classical) approximation and integrating out itinerant quasiparticles gives the following RKKY interaction between them to second order in $1/N$
\begin{eqnarray}
  \frac{\langle H_{\rm RKKY}\rangle}{N}  &=& -\frac{J_K^2}{ N^2 N_S} \sum_{\mathbf{q}, i j}\chi_{\rm cc}(\mathbf{q})  e^{i \mathbf{q} \cdot(\mathbf{r}_i-\mathbf{r}_j)} \bar{S}_{2i}\bar{S}_{2j},
  \nonumber\\
\end{eqnarray}
where $N_S$ is the number of lattice sites, $\bar{S}_{2i}$ is the expectation of the $f_2$-fermion spin on site $i$ per replica, and $\chi_{\rm cc}(\mathbf{q}) $ is the susceptibility of the conduction electrons to a $\mathbf{q}$-modulated magnetic field that is applied only to them. Details of the derivation can be found in Apps. \ref{app:gauge}-\ref{UVA} and App. \ref{sec:limit}.

Away from the $N\rightarrow\infty$ limit, it is difficult to decouple $H_N$ simultaneously in the RKKY and Kondo channels in an unbiased way and a number of different decoupling schemes exist in the literature to treat Eqs. (\ref{Hamil}) and (\ref{FermiHam}) \cite{7zsl-4497, PhysRevB.87.205107,Perkins_2007, PhysRevB.76.125101,Thomas2011, THOMAS2014}. To avoid any bias, we will begin with a variational ansatz for the ground state of $H_N$ that is exact in two limits: $N\rightarrow\infty$, where $f_2$-moments decouple entirely, and $J_K\rho \rightarrow 0$, where the coherence temperature $T_K$ is exponentially suppressed and we have a conventional RKKY ferromagnet. (In the latter limit, we take the physical dimension $d\rightarrow \infty$, so that the ordered local moments can be approximated classically. We also assume, for simplicity and without loss of generality, that the preferred order is ferromagnetic -- this allows us to work in the constant density of states approximation for the conduction band.)  

We use the following trial ground state wavefunction, introduced in our earlier work \cite{7zsl-4497},
\begin{eqnarray}\label{ansatz}
    | \Psi\rangle\! =\!\!\! \prod_{\mathbf{k\alpha\sigma \nu}} \left(u^{\nu}_{\mathbf{k}\sigma} c^{\dagger}_{\alpha\sigma}(\mathbf{k})+ v^{\nu}_{\mathbf{k}\sigma}f^{\dagger}_{1\alpha\sigma}(\mathbf{k})\right) |0\rangle \otimes  \prod_{i} | \psi_{2i} \rangle,
\end{eqnarray}
where $| \psi_{2i} \rangle$ describes the state of the $f_{2\alpha i}$-fermions on the $i$-th lattice site, $|0\rangle$ is the fermionic vacuum, and $\nu=\pm$ indexes the lower ($-$) and upper $(+)$ heavy-fermion bands. The variational parameters $\{u^{\nu}_{\mathbf{k}\alpha},v^{\nu}_{\mathbf{k}\alpha}\}$ are chosen to minimise $\langle \Psi |H_N | \Psi \rangle$ subject to the constraints on $\langle \hat{n}^f_{ai} \rangle$ and the total number of conduction electrons. This expectation value can be computed using Wick's theorem and is equal to $\langle \Psi |H_{\rm MF}| \Psi \rangle$, where $H_{\rm MF}$ is an appropriately chosen mean-field Hamiltonian 
\begin{eqnarray}
H_{\rm MF}=&\sum_{\mathbf{k} \alpha\sigma} \left( \sigma J_{M} \bar{s}  f^{\dagger}_{2 \alpha\sigma}(\mathbf{k}) f_{2\alpha\sigma}(\mathbf{k})  +\Psi_{\alpha\sigma\mathbf{k}}^{\dagger}\mathcal{H} \Psi_{\alpha\sigma\mathbf{k}} \right) 
\nonumber\\
&\;\;+ N_s N J_K \bar{V}_1^2 +\frac{1}{2}\mu N n_c N_s 
-\frac{1}{2}\bar{\lambda}_1 N N_s,
\end{eqnarray}
with the matrix $\mathcal{H}$ given by
\begin{align}
 \mathcal{H}_{}&= \begin{pmatrix}
        \epsilon_{\mathbf{k}} - \mu + J_M \sigma(\bar{S}_{1} +\bar{S}_{2}) &-J_K\bar{V}_1\\ -J_K\bar{V}_1& \bar{\lambda}_1+ \sigma J_M\bar{s}
    \end{pmatrix}.
\end{align}
$\Psi_{\alpha\sigma\mathbf{k}}^{\dagger}= \left(c_{\alpha\sigma}^{\dagger}(\mathbf{k}), f_{1 \alpha\sigma}^{\dagger}((\mathbf{k}))\right)$ and $J_{\rm M}=\frac{2}{N}J_K$ is the strength of the coupling in the magnetic exchange channel. The Lagrange multipliers $\bar{\lambda}_1$ and $\mu$ are adjusted to give the expectation values $\langle \hat{n}^{f}_{1i}\rangle =\frac{N}{2}$ and $\sum_{\sigma} \langle c^{\dagger}_{\alpha\sigma}(\mathbf{r}_i)c_{\alpha\sigma}(\mathbf{r}_i)\rangle =n_c$. Note that the constraint $\hat{n}^f_{2i}=\frac{N}{2}$ is treated exactly, which corresponds to the free fluctuations of the $\lambda_{2i}:=i\theta_i-i\phi^z_i$ gauge field and the residual $U(1)$ symmetry of the mean-field state. For magnetisation along the $z$-axis, the magnetic order parameters are thus 
\begin{eqnarray}\label{self consistent}
    \bar{S}_{ a} &=& \sum_{\sigma} \sigma \left\langle  f^{\dagger}_{a \alpha\sigma}(\mathbf{r}_i) f_{a\alpha\sigma}(\mathbf{r}_i)\right\rangle,
    \nonumber\\
    \bar{s} &=&  \sum_{\sigma} \sigma\left\langle  c^{\dagger}_{ \alpha\sigma}(\mathbf{r}_i) c_{\alpha\sigma}(\mathbf{r}_i)\right\rangle,
\end{eqnarray}
and together with $\bar{V}_1=\langle \hat{V}_{1i}\rangle$, and the constraint equations, form the system of self-consistent mean-field equations. The $|\Psi\rangle$ which minimises  $\langle \Psi |H_N| \Psi \rangle$ is the ground state of $H_{\rm MF}$. Details can be found in App. \ref{UVA}.

\section{Ground state phase diagram} \label{sec:ground state} 
\begin{figure}
\centering\includegraphics[width=0.48\textwidth]{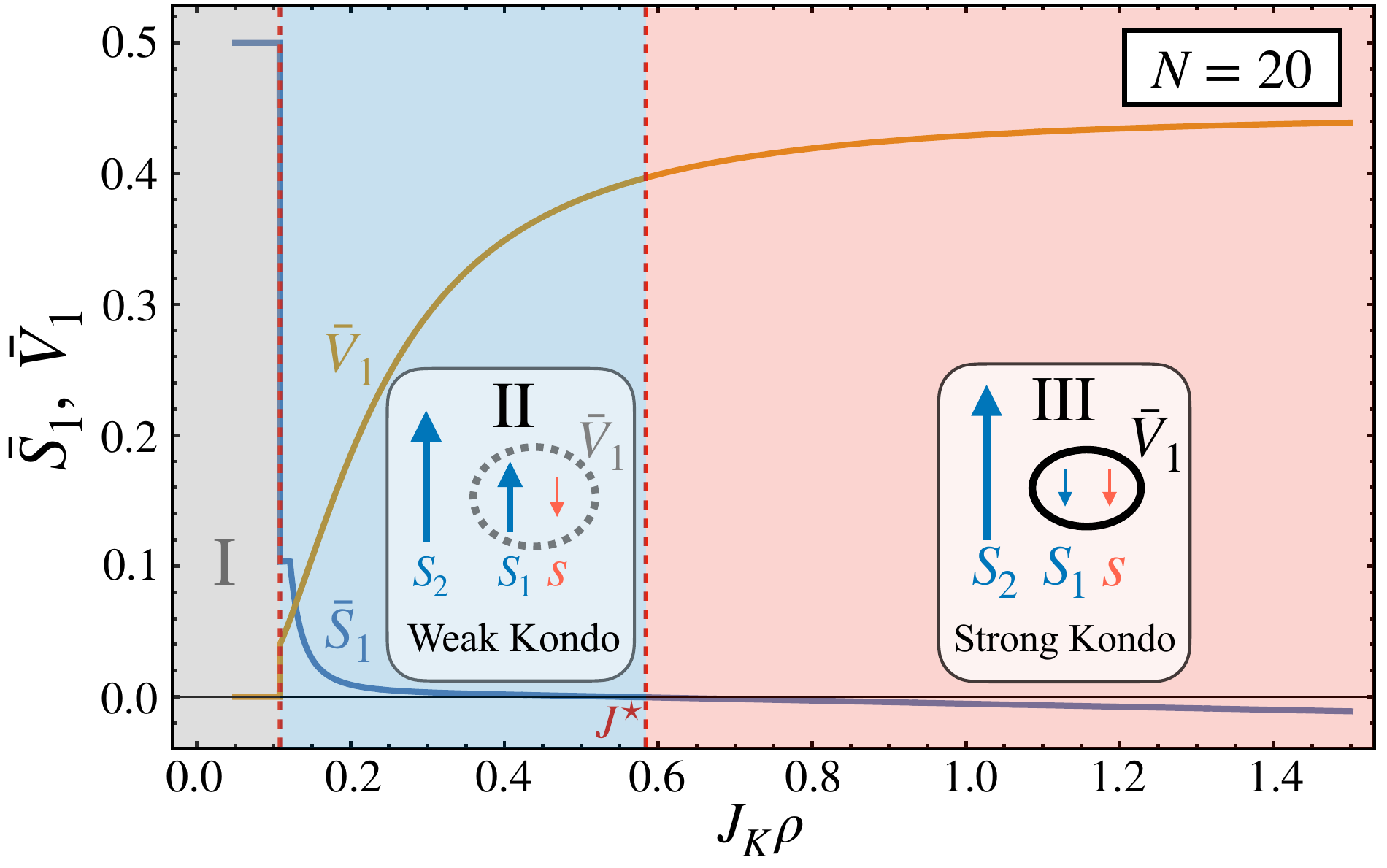}
    \caption{A plot of the groundstate order parameters $\bar{S}_1$ and $\bar{V}_1$, magnetization and hybridization of the $f_1$-fermion respectively for $N=20$. The unhybridized $f_2$-fermion always has a magnetization of $\bar{S}_2=\frac{1}{2}$. 
    The crossover between the weak and strong Kondo FM phases occurs at $ J_K\rho\sim J^\star$, defined as the point where $\bar{S}_1$ changes sign. 
    }
    \label{fig:orderparam}
\end{figure}

By solving the self-consistent mean-field equations in Eq. \ref{self consistent}, we have obtained the $T=0$ phase diagrams plotted in Fig. \ref{fig:phasedia}. The $N=20$ cut is shown in Fig. \ref{fig:orderparam}. All our numerical results are given for the representative electron filling of $n_c=0.8$.  We identify three distinct phases, with the magnetization of the unhybridized $f_2$-fermion $\bar{S}_2=1/2$ and the magnetization of conduction electrons $\bar{s}<0$ in all phases,
\begin{enumerate}[I]
    \item The normal ferromagnet (FM) phase, where there is no Kondo hybridisation ($\bar{V}_1=0$) and the spin-1 impurity is fully polarised $\bar{S}_1=\bar{S}_2=1/2$. As expected, we enter the purely magnetic phase when the Kondo energy $T_K\sim \Lambda e^{-1/(J_K \rho)}$ is lower than the RKKY energy $T_{\rm RKKY} \sim J_K^2 \rho/N^2$.
    \item The intermediate weak Kondo FM phase, where the hybridisation is nonzero but small and the magnetisation $\bar{S}_1$ is reduced but still significant. The moment of the hybridised $f_1$-fermion is positive and anti-aligned with the conduction electrons. Since, the $f_2$-moment is always anti-aligned with the conduction electrons, the total magnetisation of the impurity is between $1/2$ and 1.
    \item The large-$N$ strong Kondo ferromagnet phase, where $T_K \gg T_{\rm RKKY}$ and $\bar{S}_1$ is small and aligned with the conduction electrons.  The hybridisation is large and the total magnetisation of the impurity is less than $1/2$.
\end{enumerate}
The strong Kondo ferromagnet (phase III) is the most reliable conclusion of our work as it neighbours the $N\rightarrow\infty$ limit, where the ground state MF solution becomes exact. As expected, for any finite $N$, it becomes energetically favourable for the RKKY channel to lift the degeneracy of the $f_2$-subspace and magnetic order to form. There is a crossover from the strong to the weak Kondo FM, either as $N$ or $J_K$ fall. We mark the crossover between these two phases as the point where $\bar{S}_1$ changes sign. There is also a first order phase transition between the normal FM and the weak Kondo FM, evidenced by the jump in the order parameters seen in Fig. \ref{fig:orderparam}. The transition becomes a crossover if dynamical fluctuations are included, e.g., in DMFT \cite{Golez2013}. The transition can be understood in the Doniach picture, where the pure RKKY FM phase has the lowest energy for a sufficiently small coupling strength, $J_K\rho$. 
We also note the existence of magnetisation plateaus as $J_K\rho$ approaches the first order transition from above, as seen in Fig. \ref{fig:orderparam}. This is a consequence of one of the lower heavy-fermion bands becoming insulating and preventing the system from magnetising further. This 'spin-selective' Kondo insulator has been observed in previous studies, including DMFT \cite{PhysRevLett.108.086402}. 

In the $N\rightarrow$ limit, the order parameters can be analytically computed and are given in App. \ref{sec:limit}. In particular, the magnetizations $\bar{S}_1=-\frac{1}{2}J_{\rm M} \chi_{\rm cf}(\mathbf{0})$ and $\bar{s}=-\frac{1}{2}J_{\rm M} \chi_{\rm cc}(\mathbf{0})$ are both of order $1/N$. $\chi_{\rm cf}(\mathbf{0})$ is the susceptibility of the hybridized $f_1$-moments in the $N\rightarrow \infty$ limit when the magnetic field is only applied to conduction electrons. The magnetization of the hybridized $f_1$-fermion, $\bar{S_1}$, is aligned with the magnetization of the conduction electron, $\bar{s}$, in the strong Kondo FM phase because $\chi_{\rm cf} (\mathbf{0})>0$. The sign of the mixed susceptibility can be understood as follows. The Fermi liquid quasiparticle excitations are thought to be a superposition of the fractionalised $f_1$-moments and conduction electrons with the same spin quantum number (see Ref. \cite{PhysRevB.76.153101} for example), such that a magnetic field applied to only one of the two species lowers the whole majority-spin quasiparticle band and increases the occupation of the other species of the same spin. 
We note that this is not seen in the insulating limit ($n_c=1$), where, in the strong-coupling limit, the Kondo lattice can be viewed as an array of singlets (and decoupled $f_2$-moments). The insulator is incompressible so that a weak field applied to one of the two species can only lead to an asymmetry of the singlet eigenstates without changing their occupation, and as expected $\chi_{\rm cf} (\mathbf{0})<0$ in this case.

 One can estimate the value of the magnetizations in the physical $N=2$ limit by simply extrapolating the above results of the expansion around the $N\rightarrow\infty$ limit to first order in $1/N$. This is equivalent to the approach taken in Refs. \cite{PhysRevB.87.205107,7zsl-4497}.
The mean-field decouplings used in Refs. \cite{Perkins_2007, PhysRevB.76.125101,Thomas2011, THOMAS2014}, on the other hand, largely correspond to the results of our variational approach for $N=2$. Here, we are always in the weak Kondo FM phase, where the magnetic exchange channel dominates over the hybridization channel in setting the relative signs of the magnetizations: $\bar{S}_{1,2}$ have the same sign and are both antialigned with respect to the magnetization of the conduction electrons $\bar{s}$. We note a couple of differences between the decouplings used in these references and our theory for $N=2$. Firstly, the same value of the chemical potential was taken for the unhybridized $f$-fermion as for the hybridized one, which would correspond to setting $\bar{\lambda}_1=\bar{\lambda}_2$ in our theory (we instead always set the chemical potential of the unhybridized $f$-fermion to zero). This should have no effect on the results for small enough coupling $J_K\rho$, where the magnetization of conduction electrons is large enough by comparison with $\bar{\lambda}_1/J_M$, resulting in the unhybridized $f$-fermion state being singly occupied and fully magnetized. However, in Refs. \cite{Perkins_2007}-\cite{PhysRevB.76.125101} the magnetic order collapses for sufficiently large coupling, which we do not find for $N=2$, presumably because of the different treatment of the chemical potential for the unhybridized $f$-moment.   Secondly, in some of Refs. \cite{Perkins_2007, PhysRevB.76.125101,Thomas2011, THOMAS2014} the entire Kondo exchange is decoupled in the hybridization channel, while in others, the exchange along the direction of magnetic order is not decoupled in this channel. Our variational results for $N=2$ are equivalent to a mean-field decoupling of the entire exchange in the Kondo channel. We note that the differences between these different approaches (apart from a rescaled Kondo coupling) should be small if the hybridizations of the up and down fermions are similar. The results of Ref. \cite{Perkins_2007} show that this is indeed the case. 

The above discussion leaves the question of which mean-field decoupling is closer to describing physical systems: (i) performing the $1/N$ expansion to first order and extrapolating to $N=2$; (ii) setting $N=2$ from the outset. They certainly result in a different structure of  coexistent Kondo hybridization and magnetic order, with only (i) leading to stabilization of hard-direction order \cite{7zsl-4497}, which is why it is important to address this question. Approach (i) is certainly more controlled, as fluctuation corrections are small in the $N\rightarrow\infty$ limit. It is similar to performing a $1/S$ expansion and extrapolating it to $S=1/2$, e.g., as in the gMFT treatment of quantum spin ice \cite{PhysRevB.95.134439}. Extrapolating such expansions has often even given good quantitative agreement and can be understood as the control parameter $S$ or $N$ flowing to infinity in an RG sense. We thus expect approach (i) to reflect the physical system more closely if it flows to the fixed-gauge saddle point in the RG sense and magnetic interactions can be considered perturbatively. Conversely, we expect approach (ii), and the resulting weak Kondo magnetic phase, to be a closer reflection of the physical situation if the magnetic and Kondo interactions are equally dominant, e.g., close to the transition to the FM phase with no hybridization.

\section{Conclusion}
In summary, we have extended the standard large-$N$ Read-Newns theory to the case of the spin-$1$ underscreened Kondo lattice (UKL) and found the fluctuations around the $N\rightarrow \infty$ saddle point to be just as small as in the standard spin-$1/2$ Kondo lattice, which is manifested by the same rate of decoherence of the hybridization field with time. In particular, we derived the residual ferromagnetic interaction between the unhybridized moment and the heavy-fermion Fermi liquid, which is caused by transverse, long-wavelength fluctuations of the hybridization field, and found it to decay logarithmically with temperature. 

Unlike for the case of a single spin-1 Kondo impurity, for a lattice of impurities, the coexistence of Kondo hybridization and magnetic order is possible, and RKKY interactions between the residual moments also need to be considered. These are generated by short-wavelength transverse fluctuations of the hybridization field, and we compute the analytic form in the $N\rightarrow\infty$ limit.
Employing a variational ansatz, exact for $N\rightarrow \infty$ and $J_K\rho\rightarrow0$ limits (the latter in sufficiently high dimension), we showed that it is always energetically favourable for the unscreened residual moments to order via the RKKY magnetic channel and that the coexistence of interactions in the RKKY and Kondo hybridisation channels leads to two distinct phases with a crossover between them. The weak Kondo regime where the hybridised $f$-moment retains some local character and aligns with the unhybridised $f$-moment, and the strong Kondo regime where the hybridised $f$-moment is fully injected into the Fermi sea and its magnetization antialigns with the unhybridised $f$-moment, but aligns with that of the conduction electrons.

Our variational ansatz unites some of the previously proposed MF decouplings for the spin-1 UKL, which, within our variational approach, correspond to: (i) taking the $N\rightarrow\infty$ and extrapolating the results to the physical $N=2$ limit; (ii) setting $N=2$ from the outset. The structure of the resulting coexistent ground states is different, weak Kondo FM phase in case (i) and the strong Kondo FM phase in case (ii). We show that while fluctuations are controlled in the first case, where Kondo hybridization dominates over magnetic exchange, the latter might better describe the coexistent phase in the vicinity of the transition to the normal FM magnet.

\bibliography{apssamp}

\begin{thebibliography}{34}%
\makeatletter
\providecommand \@ifxundefined [1]{%
 \@ifx{#1\undefined}
}%
\providecommand \@ifnum [1]{%
 \ifnum #1\expandafter \@firstoftwo
 \else \expandafter \@secondoftwo
 \fi
}%
\providecommand \@ifx [1]{%
 \ifx #1\expandafter \@firstoftwo
 \else \expandafter \@secondoftwo
 \fi
}%
\providecommand \natexlab [1]{#1}%
\providecommand \enquote  [1]{``#1''}%
\providecommand \bibnamefont  [1]{#1}%
\providecommand \bibfnamefont [1]{#1}%
\providecommand \citenamefont [1]{#1}%
\providecommand \href@noop [0]{\@secondoftwo}%
\providecommand \href [0]{\begingroup \@sanitize@url \@href}%
\providecommand \@href[1]{\@@startlink{#1}\@@href}%
\providecommand \@@href[1]{\endgroup#1\@@endlink}%
\providecommand \@sanitize@url [0]{\catcode `\\12\catcode `\$12\catcode `\&12\catcode `\#12\catcode `\^12\catcode `\_12\catcode `\%12\relax}%
\providecommand \@@startlink[1]{}%
\providecommand \@@endlink[0]{}%
\providecommand \url  [0]{\begingroup\@sanitize@url \@url }%
\providecommand \@url [1]{\endgroup\@href {#1}{\urlprefix }}%
\providecommand \urlprefix  [0]{URL }%
\providecommand \Eprint [0]{\href }%
\providecommand \doibase [0]{http://dx.doi.org/}%
\providecommand \selectlanguage [0]{\@gobble}%
\providecommand \bibinfo  [0]{\@secondoftwo}%
\providecommand \bibfield  [0]{\@secondoftwo}%
\providecommand \translation [1]{[#1]}%
\providecommand \BibitemOpen [0]{}%
\providecommand \bibitemStop [0]{}%
\providecommand \bibitemNoStop [0]{.\EOS\space}%
\providecommand \EOS [0]{\spacefactor3000\relax}%
\providecommand \BibitemShut  [1]{\csname bibitem#1\endcsname}%
\let\auto@bib@innerbib\@empty
\bibitem [{\citenamefont {Zapf}\ \emph {et~al.}(2001)\citenamefont {Zapf}, \citenamefont {Freeman}, \citenamefont {Bauer}, \citenamefont {Petricka}, \citenamefont {Sirvent}, \citenamefont {Frederick}, \citenamefont {Dickey},\ and\ \citenamefont {Maple}}]{Ce_super}%
  \BibitemOpen
  \bibfield  {author} {\bibinfo {author} {\bibfnamefont {V.~S.}\ \bibnamefont {Zapf}}, \bibinfo {author} {\bibfnamefont {E.~J.}\ \bibnamefont {Freeman}}, \bibinfo {author} {\bibfnamefont {E.~D.}\ \bibnamefont {Bauer}}, \bibinfo {author} {\bibfnamefont {J.}~\bibnamefont {Petricka}}, \bibinfo {author} {\bibfnamefont {C.}~\bibnamefont {Sirvent}}, \bibinfo {author} {\bibfnamefont {N.~A.}\ \bibnamefont {Frederick}}, \bibinfo {author} {\bibfnamefont {R.~P.}\ \bibnamefont {Dickey}}, \ and\ \bibinfo {author} {\bibfnamefont {M.~B.}\ \bibnamefont {Maple}},\ }\href {\doibase 10.1103/PhysRevB.65.014506} {\bibfield  {journal} {\bibinfo  {journal} {Phys. Rev. B}\ }\textbf {\bibinfo {volume} {65}},\ \bibinfo {pages} {014506} (\bibinfo {year} {2001})}\BibitemShut {NoStop}%
\bibitem [{\citenamefont {Aoki}\ \emph {et~al.}(2001)\citenamefont {Aoki}, \citenamefont {Huxley}, \citenamefont {Ressouche}, \citenamefont {Braithwaite}, \citenamefont {Flouquet}, \citenamefont {Brison}, \citenamefont {Lhotel},\ and\ \citenamefont {Paulsen}}]{URhGe}%
  \BibitemOpen
  \bibfield  {author} {\bibinfo {author} {\bibfnamefont {D.}~\bibnamefont {Aoki}}, \bibinfo {author} {\bibfnamefont {A.}~\bibnamefont {Huxley}}, \bibinfo {author} {\bibfnamefont {E.}~\bibnamefont {Ressouche}}, \bibinfo {author} {\bibfnamefont {D.}~\bibnamefont {Braithwaite}}, \bibinfo {author} {\bibfnamefont {J.}~\bibnamefont {Flouquet}}, \bibinfo {author} {\bibfnamefont {J.-P.}\ \bibnamefont {Brison}}, \bibinfo {author} {\bibfnamefont {E.}~\bibnamefont {Lhotel}}, \ and\ \bibinfo {author} {\bibfnamefont {C.}~\bibnamefont {Paulsen}},\ }\href {\doibase 10.1038/35098048} {\bibfield  {journal} {\bibinfo  {journal} {Nature}\ }\textbf {\bibinfo {volume} {413}},\ \bibinfo {pages} {613} (\bibinfo {year} {2001})}\BibitemShut {NoStop}%
\bibitem [{\citenamefont {Ran}\ \emph {et~al.}(2019)\citenamefont {Ran}, \citenamefont {Eckberg}, \citenamefont {Ding}, \citenamefont {Furukawa}, \citenamefont {Metz}, \citenamefont {Saha}, \citenamefont {Liu}, \citenamefont {Zic}, \citenamefont {Kim}, \citenamefont {Paglione},\ and\ \citenamefont {Butch}}]{UTe2}%
  \BibitemOpen
  \bibfield  {author} {\bibinfo {author} {\bibfnamefont {S.}~\bibnamefont {Ran}}, \bibinfo {author} {\bibfnamefont {C.}~\bibnamefont {Eckberg}}, \bibinfo {author} {\bibfnamefont {Q.-P.}\ \bibnamefont {Ding}}, \bibinfo {author} {\bibfnamefont {Y.}~\bibnamefont {Furukawa}}, \bibinfo {author} {\bibfnamefont {T.}~\bibnamefont {Metz}}, \bibinfo {author} {\bibfnamefont {S.~R.}\ \bibnamefont {Saha}}, \bibinfo {author} {\bibfnamefont {I.-L.}\ \bibnamefont {Liu}}, \bibinfo {author} {\bibfnamefont {M.}~\bibnamefont {Zic}}, \bibinfo {author} {\bibfnamefont {H.}~\bibnamefont {Kim}}, \bibinfo {author} {\bibfnamefont {J.}~\bibnamefont {Paglione}}, \ and\ \bibinfo {author} {\bibfnamefont {N.~P.}\ \bibnamefont {Butch}},\ }\href {\doibase 10.1126/science.aav8645} {\bibfield  {journal} {\bibinfo  {journal} {Science}\ }\textbf {\bibinfo {volume} {365}},\ \bibinfo {pages} {684} (\bibinfo {year} {2019})}\BibitemShut {NoStop}%
\bibitem [{\citenamefont {Shen}\ \emph {et~al.}(2020)\citenamefont {Shen}, \citenamefont {Zhang}, \citenamefont {Komijani}, \citenamefont {Nicklas}, \citenamefont {Borth}, \citenamefont {Wang}, \citenamefont {Chen}, \citenamefont {Nie}, \citenamefont {Li}, \citenamefont {Lu}, \citenamefont {Lee}, \citenamefont {Smidman}, \citenamefont {Steglich}, \citenamefont {Coleman},\ and\ \citenamefont {Yuan}}]{Strange_Metal}%
  \BibitemOpen
  \bibfield  {author} {\bibinfo {author} {\bibfnamefont {B.}~\bibnamefont {Shen}}, \bibinfo {author} {\bibfnamefont {Y.}~\bibnamefont {Zhang}}, \bibinfo {author} {\bibfnamefont {Y.}~\bibnamefont {Komijani}}, \bibinfo {author} {\bibfnamefont {M.}~\bibnamefont {Nicklas}}, \bibinfo {author} {\bibfnamefont {R.}~\bibnamefont {Borth}}, \bibinfo {author} {\bibfnamefont {A.}~\bibnamefont {Wang}}, \bibinfo {author} {\bibfnamefont {Y.}~\bibnamefont {Chen}}, \bibinfo {author} {\bibfnamefont {Z.}~\bibnamefont {Nie}}, \bibinfo {author} {\bibfnamefont {R.}~\bibnamefont {Li}}, \bibinfo {author} {\bibfnamefont {X.}~\bibnamefont {Lu}}, \bibinfo {author} {\bibfnamefont {H.}~\bibnamefont {Lee}}, \bibinfo {author} {\bibfnamefont {M.}~\bibnamefont {Smidman}}, \bibinfo {author} {\bibfnamefont {F.}~\bibnamefont {Steglich}}, \bibinfo {author} {\bibfnamefont {P.}~\bibnamefont {Coleman}}, \ and\ \bibinfo {author} {\bibfnamefont {H.}~\bibnamefont {Yuan}},\ }\href {\doibase 10.1038/s41586-020-2052-z} {\bibfield  {journal} {\bibinfo
  {journal} {Nature}\ }\textbf {\bibinfo {volume} {579}},\ \bibinfo {pages} {51} (\bibinfo {year} {2020})}\BibitemShut {NoStop}%
\bibitem [{\citenamefont {Lévy}\ \emph {et~al.}(2005)\citenamefont {Lévy}, \citenamefont {Sheikin}, \citenamefont {Grenier},\ and\ \citenamefont {Huxley}}]{URhGe_metamagnetic}%
  \BibitemOpen
  \bibfield  {author} {\bibinfo {author} {\bibfnamefont {F.}~\bibnamefont {Lévy}}, \bibinfo {author} {\bibfnamefont {I.}~\bibnamefont {Sheikin}}, \bibinfo {author} {\bibfnamefont {B.}~\bibnamefont {Grenier}}, \ and\ \bibinfo {author} {\bibfnamefont {A.~D.}\ \bibnamefont {Huxley}},\ }\href {\doibase 10.1126/science.1115498} {\bibfield  {journal} {\bibinfo  {journal} {Science}\ }\textbf {\bibinfo {volume} {309}},\ \bibinfo {pages} {1343} (\bibinfo {year} {2005})}\BibitemShut {NoStop}%
\bibitem [{\citenamefont {Hafner}\ \emph {et~al.}(2019)\citenamefont {Hafner}, \citenamefont {Rai}, \citenamefont {Banda}, \citenamefont {Kliemt}, \citenamefont {Krellner}, \citenamefont {Sichelschmidt}, \citenamefont {Morosan}, \citenamefont {Geibel},\ and\ \citenamefont {Brando}}]{Brando}%
  \BibitemOpen
  \bibfield  {author} {\bibinfo {author} {\bibfnamefont {D.}~\bibnamefont {Hafner}}, \bibinfo {author} {\bibfnamefont {B.~K.}\ \bibnamefont {Rai}}, \bibinfo {author} {\bibfnamefont {J.}~\bibnamefont {Banda}}, \bibinfo {author} {\bibfnamefont {K.}~\bibnamefont {Kliemt}}, \bibinfo {author} {\bibfnamefont {C.}~\bibnamefont {Krellner}}, \bibinfo {author} {\bibfnamefont {J.}~\bibnamefont {Sichelschmidt}}, \bibinfo {author} {\bibfnamefont {E.}~\bibnamefont {Morosan}}, \bibinfo {author} {\bibfnamefont {C.}~\bibnamefont {Geibel}}, \ and\ \bibinfo {author} {\bibfnamefont {M.}~\bibnamefont {Brando}},\ }\href {\doibase 10.1103/PhysRevB.99.201109} {\bibfield  {journal} {\bibinfo  {journal} {Phys. Rev. B}\ }\textbf {\bibinfo {volume} {99}},\ \bibinfo {pages} {201109} (\bibinfo {year} {2019})}\BibitemShut {NoStop}%
\bibitem [{\citenamefont {Scott}\ and\ \citenamefont {Kwasigroch}(2025)}]{7zsl-4497}%
  \BibitemOpen
  \bibfield  {author} {\bibinfo {author} {\bibfnamefont {E.}~\bibnamefont {Scott}}\ and\ \bibinfo {author} {\bibfnamefont {M.}~\bibnamefont {Kwasigroch}},\ }\href {\doibase 10.1103/7zsl-4497} {\bibfield  {journal} {\bibinfo  {journal} {Phys. Rev. Res.}\ }\textbf {\bibinfo {volume} {7}},\ \bibinfo {pages} {043225} (\bibinfo {year} {2025})}\BibitemShut {NoStop}%
\bibitem [{\citenamefont {Ji}\ \emph {et~al.}(2022)\citenamefont {Ji}, \citenamefont {Luo}, \citenamefont {Chen}, \citenamefont {Feng}, \citenamefont {Hao}, \citenamefont {Liu}, \citenamefont {Zhang}, \citenamefont {Liu}, \citenamefont {Wang}, \citenamefont {Tan},\ and\ \citenamefont {Lai}}]{UAs2}%
  \BibitemOpen
  \bibfield  {author} {\bibinfo {author} {\bibfnamefont {X.}~\bibnamefont {Ji}}, \bibinfo {author} {\bibfnamefont {X.}~\bibnamefont {Luo}}, \bibinfo {author} {\bibfnamefont {Q.}~\bibnamefont {Chen}}, \bibinfo {author} {\bibfnamefont {W.}~\bibnamefont {Feng}}, \bibinfo {author} {\bibfnamefont {Q.}~\bibnamefont {Hao}}, \bibinfo {author} {\bibfnamefont {Q.}~\bibnamefont {Liu}}, \bibinfo {author} {\bibfnamefont {Y.}~\bibnamefont {Zhang}}, \bibinfo {author} {\bibfnamefont {Y.}~\bibnamefont {Liu}}, \bibinfo {author} {\bibfnamefont {X.}~\bibnamefont {Wang}}, \bibinfo {author} {\bibfnamefont {S.}~\bibnamefont {Tan}}, \ and\ \bibinfo {author} {\bibfnamefont {X.}~\bibnamefont {Lai}},\ }\href {\doibase 10.1103/PhysRevB.106.125120} {\bibfield  {journal} {\bibinfo  {journal} {Phys. Rev. B}\ }\textbf {\bibinfo {volume} {106}},\ \bibinfo {pages} {125120} (\bibinfo {year} {2022})}\BibitemShut {NoStop}%
\bibitem [{\citenamefont {Auerbach}\ and\ \citenamefont {Levin}(1986)}]{PhysRevLett.57.877}%
  \BibitemOpen
  \bibfield  {author} {\bibinfo {author} {\bibfnamefont {A.}~\bibnamefont {Auerbach}}\ and\ \bibinfo {author} {\bibfnamefont {K.}~\bibnamefont {Levin}},\ }\href@noop {} {\bibfield  {journal} {\bibinfo  {journal} {Phys. Rev. Lett.}\ }\textbf {\bibinfo {volume} {57}},\ \bibinfo {pages} {877} (\bibinfo {year} {1986})}\BibitemShut {NoStop}%
\bibitem [{\citenamefont {Read}\ and\ \citenamefont {Newns}(1983)}]{NRead_1983}%
  \BibitemOpen
  \bibfield  {author} {\bibinfo {author} {\bibfnamefont {N.}~\bibnamefont {Read}}\ and\ \bibinfo {author} {\bibfnamefont {D.~M.}\ \bibnamefont {Newns}},\ }\href {\doibase 10.1088/0022-3719/16/17/014} {\bibfield  {journal} {\bibinfo  {journal} {Journal of Physics C: Solid State Physics}\ }\textbf {\bibinfo {volume} {16}},\ \bibinfo {pages} {3273} (\bibinfo {year} {1983})}\BibitemShut {NoStop}%
\bibitem [{\citenamefont {McCulloch}\ \emph {et~al.}(2002)\citenamefont {McCulloch}, \citenamefont {Juozapavicius}, \citenamefont {Rosengren},\ and\ \citenamefont {Gulacsi}}]{McCulloch2002}%
  \BibitemOpen
  \bibfield  {author} {\bibinfo {author} {\bibfnamefont {I.~P.}\ \bibnamefont {McCulloch}}, \bibinfo {author} {\bibfnamefont {A.}~\bibnamefont {Juozapavicius}}, \bibinfo {author} {\bibfnamefont {A.}~\bibnamefont {Rosengren}}, \ and\ \bibinfo {author} {\bibfnamefont {M.}~\bibnamefont {Gulacsi}},\ }\href {\doibase 10.1103/PhysRevB.65.052410} {\bibfield  {journal} {\bibinfo  {journal} {Phys. Rev. B}\ }\textbf {\bibinfo {volume} {65}},\ \bibinfo {pages} {052410} (\bibinfo {year} {2002})}\BibitemShut {NoStop}%
\bibitem [{\citenamefont {Masui}\ and\ \citenamefont {Totsuka}(2022)}]{Masui2022}%
  \BibitemOpen
  \bibfield  {author} {\bibinfo {author} {\bibfnamefont {R.}~\bibnamefont {Masui}}\ and\ \bibinfo {author} {\bibfnamefont {K.}~\bibnamefont {Totsuka}},\ }\href {\doibase 10.1103/PhysRevB.106.014411} {\bibfield  {journal} {\bibinfo  {journal} {Phys. Rev. B}\ }\textbf {\bibinfo {volume} {106}},\ \bibinfo {pages} {014411} (\bibinfo {year} {2022})}\BibitemShut {NoStop}%
\bibitem [{\citenamefont {Nakatsuji}\ \emph {et~al.}(2004)\citenamefont {Nakatsuji}, \citenamefont {Pines},\ and\ \citenamefont {Fisk}}]{PhysRevLett.92.016401}%
  \BibitemOpen
  \bibfield  {author} {\bibinfo {author} {\bibfnamefont {S.}~\bibnamefont {Nakatsuji}}, \bibinfo {author} {\bibfnamefont {D.}~\bibnamefont {Pines}}, \ and\ \bibinfo {author} {\bibfnamefont {Z.}~\bibnamefont {Fisk}},\ }\href {\doibase 10.1103/PhysRevLett.92.016401} {\bibfield  {journal} {\bibinfo  {journal} {Phys. Rev. Lett.}\ }\textbf {\bibinfo {volume} {92}},\ \bibinfo {pages} {016401} (\bibinfo {year} {2004})}\BibitemShut {NoStop}%
\bibitem [{\citenamefont {Ramires}\ and\ \citenamefont {Coleman}(2016)}]{PhysRevB.93.035120}%
  \BibitemOpen
  \bibfield  {author} {\bibinfo {author} {\bibfnamefont {A.}~\bibnamefont {Ramires}}\ and\ \bibinfo {author} {\bibfnamefont {P.}~\bibnamefont {Coleman}},\ }\href {\doibase 10.1103/PhysRevB.93.035120} {\bibfield  {journal} {\bibinfo  {journal} {Phys. Rev. B}\ }\textbf {\bibinfo {volume} {93}},\ \bibinfo {pages} {035120} (\bibinfo {year} {2016})}\BibitemShut {NoStop}%
\bibitem [{\citenamefont {Coleman}\ \emph {et~al.}(2000)\citenamefont {Coleman}, \citenamefont {P\'epin},\ and\ \citenamefont {Tsvelik}}]{PhysRevB.62.3852}%
  \BibitemOpen
  \bibfield  {author} {\bibinfo {author} {\bibfnamefont {P.}~\bibnamefont {Coleman}}, \bibinfo {author} {\bibfnamefont {C.}~\bibnamefont {P\'epin}}, \ and\ \bibinfo {author} {\bibfnamefont {A.~M.}\ \bibnamefont {Tsvelik}},\ }\href {\doibase 10.1103/PhysRevB.62.3852} {\bibfield  {journal} {\bibinfo  {journal} {Phys. Rev. B}\ }\textbf {\bibinfo {volume} {62}},\ \bibinfo {pages} {3852} (\bibinfo {year} {2000})}\BibitemShut {NoStop}%
\bibitem [{\citenamefont {Raczkowski}\ and\ \citenamefont {Assaad}(2020)}]{PhysRevResearch.2.013276}%
  \BibitemOpen
  \bibfield  {author} {\bibinfo {author} {\bibfnamefont {M.}~\bibnamefont {Raczkowski}}\ and\ \bibinfo {author} {\bibfnamefont {F.~F.}\ \bibnamefont {Assaad}},\ }\href {\doibase 10.1103/PhysRevResearch.2.013276} {\bibfield  {journal} {\bibinfo  {journal} {Phys. Rev. Res.}\ }\textbf {\bibinfo {volume} {2}},\ \bibinfo {pages} {013276} (\bibinfo {year} {2020})}\BibitemShut {NoStop}%
\bibitem [{\citenamefont {Perkins}\ \emph {et~al.}(2007{\natexlab{a}})\citenamefont {Perkins}, \citenamefont {Iglesias}, \citenamefont {Núñez-Regueiro},\ and\ \citenamefont {Coqblin}}]{Perkins_2007}%
  \BibitemOpen
  \bibfield  {author} {\bibinfo {author} {\bibfnamefont {N.~B.}\ \bibnamefont {Perkins}}, \bibinfo {author} {\bibfnamefont {J.~R.}\ \bibnamefont {Iglesias}}, \bibinfo {author} {\bibfnamefont {M.~D.}\ \bibnamefont {Núñez-Regueiro}}, \ and\ \bibinfo {author} {\bibfnamefont {B.}~\bibnamefont {Coqblin}},\ }\href {\doibase 10.1209/0295-5075/79/57006} {\bibfield  {journal} {\bibinfo  {journal} {Europhysics Letters}\ }\textbf {\bibinfo {volume} {79}},\ \bibinfo {pages} {57006} (\bibinfo {year} {2007}{\natexlab{a}})}\BibitemShut {NoStop}%
\bibitem [{\citenamefont {Ko}\ \emph {et~al.}(2013)\citenamefont {Ko}, \citenamefont {Jiang}, \citenamefont {Rau},\ and\ \citenamefont {Balents}}]{PhysRevB.87.205107}%
  \BibitemOpen
  \bibfield  {author} {\bibinfo {author} {\bibfnamefont {W.-H.}\ \bibnamefont {Ko}}, \bibinfo {author} {\bibfnamefont {H.-C.}\ \bibnamefont {Jiang}}, \bibinfo {author} {\bibfnamefont {J.~G.}\ \bibnamefont {Rau}}, \ and\ \bibinfo {author} {\bibfnamefont {L.}~\bibnamefont {Balents}},\ }\href {\doibase 10.1103/PhysRevB.87.205107} {\bibfield  {journal} {\bibinfo  {journal} {Phys. Rev. B}\ }\textbf {\bibinfo {volume} {87}},\ \bibinfo {pages} {205107} (\bibinfo {year} {2013})}\BibitemShut {NoStop}%
\bibitem [{\citenamefont {Perkins}\ \emph {et~al.}(2007{\natexlab{b}})\citenamefont {Perkins}, \citenamefont {N\'u\~nez Regueiro}, \citenamefont {Coqblin},\ and\ \citenamefont {Iglesias}}]{PhysRevB.76.125101}%
  \BibitemOpen
  \bibfield  {author} {\bibinfo {author} {\bibfnamefont {N.~B.}\ \bibnamefont {Perkins}}, \bibinfo {author} {\bibfnamefont {M.~D.}\ \bibnamefont {N\'u\~nez Regueiro}}, \bibinfo {author} {\bibfnamefont {B.}~\bibnamefont {Coqblin}}, \ and\ \bibinfo {author} {\bibfnamefont {J.~R.}\ \bibnamefont {Iglesias}},\ }\href {\doibase 10.1103/PhysRevB.76.125101} {\bibfield  {journal} {\bibinfo  {journal} {Phys. Rev. B}\ }\textbf {\bibinfo {volume} {76}},\ \bibinfo {pages} {125101} (\bibinfo {year} {2007}{\natexlab{b}})}\BibitemShut {NoStop}%
\bibitem [{\citenamefont {Thomas}\ \emph {et~al.}(2011)\citenamefont {Thomas}, \citenamefont {da~Rosa Sim\~oes}, \citenamefont {Iglesias}, \citenamefont {Lacroix}, \citenamefont {Perkins},\ and\ \citenamefont {Coqblin}}]{Thomas2011}%
  \BibitemOpen
  \bibfield  {author} {\bibinfo {author} {\bibfnamefont {C.}~\bibnamefont {Thomas}}, \bibinfo {author} {\bibfnamefont {A.~S.}\ \bibnamefont {da~Rosa Sim\~oes}}, \bibinfo {author} {\bibfnamefont {J.~R.}\ \bibnamefont {Iglesias}}, \bibinfo {author} {\bibfnamefont {C.}~\bibnamefont {Lacroix}}, \bibinfo {author} {\bibfnamefont {N.~B.}\ \bibnamefont {Perkins}}, \ and\ \bibinfo {author} {\bibfnamefont {B.}~\bibnamefont {Coqblin}},\ }\href {\doibase 10.1103/PhysRevB.83.014415} {\bibfield  {journal} {\bibinfo  {journal} {Phys. Rev. B}\ }\textbf {\bibinfo {volume} {83}},\ \bibinfo {pages} {014415} (\bibinfo {year} {2011})}\BibitemShut {NoStop}%
\bibitem [{\citenamefont {Thomas}\ \emph {et~al.}(2014)\citenamefont {Thomas}, \citenamefont {{da Rosa Simões}}, \citenamefont {Lacroix}, \citenamefont {Iglesias},\ and\ \citenamefont {Coqblin}}]{THOMAS2014}%
  \BibitemOpen
  \bibfield  {author} {\bibinfo {author} {\bibfnamefont {C.}~\bibnamefont {Thomas}}, \bibinfo {author} {\bibfnamefont {A.~S.}\ \bibnamefont {{da Rosa Simões}}}, \bibinfo {author} {\bibfnamefont {C.}~\bibnamefont {Lacroix}}, \bibinfo {author} {\bibfnamefont {J.~R.}\ \bibnamefont {Iglesias}}, \ and\ \bibinfo {author} {\bibfnamefont {B.}~\bibnamefont {Coqblin}},\ }\href {\doibase https://doi.org/10.1016/j.jmmm.2014.07.028} {\bibfield  {journal} {\bibinfo  {journal} {Journal of Magnetism and Magnetic Materials}\ }\textbf {\bibinfo {volume} {372}},\ \bibinfo {pages} {247} (\bibinfo {year} {2014})}\BibitemShut {NoStop}%
\bibitem [{\citenamefont {Gole\ifmmode~\check{z}\else \v{z}\fi{}}\ and\ \citenamefont {\ifmmode~\check{Z}\else \v{Z}\fi{}itko}(2013)}]{Golez2013}%
  \BibitemOpen
  \bibfield  {author} {\bibinfo {author} {\bibfnamefont {D.}~\bibnamefont {Gole\ifmmode~\check{z}\else \v{z}\fi{}}}\ and\ \bibinfo {author} {\bibfnamefont {R.}~\bibnamefont {\ifmmode~\check{Z}\else \v{Z}\fi{}itko}},\ }\href {\doibase 10.1103/PhysRevB.88.054431} {\bibfield  {journal} {\bibinfo  {journal} {Phys. Rev. B}\ }\textbf {\bibinfo {volume} {88}},\ \bibinfo {pages} {054431} (\bibinfo {year} {2013})}\BibitemShut {NoStop}%
\bibitem [{\citenamefont {Read}(1985)}]{NRead_1985}%
  \BibitemOpen
  \bibfield  {author} {\bibinfo {author} {\bibfnamefont {N.}~\bibnamefont {Read}},\ }\href {\doibase 10.1088/0022-3719/18/13/012} {\bibfield  {journal} {\bibinfo  {journal} {Journal of Physics C: Solid State Physics}\ }\textbf {\bibinfo {volume} {18}},\ \bibinfo {pages} {2651} (\bibinfo {year} {1985})}\BibitemShut {NoStop}%
\bibitem [{\citenamefont {Coleman}(1983)}]{PhysRevB.28.5255}%
  \BibitemOpen
  \bibfield  {author} {\bibinfo {author} {\bibfnamefont {P.}~\bibnamefont {Coleman}},\ }\href {\doibase 10.1103/PhysRevB.28.5255} {\bibfield  {journal} {\bibinfo  {journal} {Phys. Rev. B}\ }\textbf {\bibinfo {volume} {28}},\ \bibinfo {pages} {5255} (\bibinfo {year} {1983})}\BibitemShut {NoStop}%
\bibitem [{\citenamefont {Wugalter}\ \emph {et~al.}(2020)\citenamefont {Wugalter}, \citenamefont {Komijani},\ and\ \citenamefont {Coleman}}]{PhysRevB.101.075133}%
  \BibitemOpen
  \bibfield  {author} {\bibinfo {author} {\bibfnamefont {A.}~\bibnamefont {Wugalter}}, \bibinfo {author} {\bibfnamefont {Y.}~\bibnamefont {Komijani}}, \ and\ \bibinfo {author} {\bibfnamefont {P.}~\bibnamefont {Coleman}},\ }\href {\doibase 10.1103/PhysRevB.101.075133} {\bibfield  {journal} {\bibinfo  {journal} {Phys. Rev. B}\ }\textbf {\bibinfo {volume} {101}},\ \bibinfo {pages} {075133} (\bibinfo {year} {2020})}\BibitemShut {NoStop}%
\bibitem [{\citenamefont {Coleman}(2015)}]{Coleman_2015}%
  \BibitemOpen
  \bibfield  {author} {\bibinfo {author} {\bibfnamefont {P.}~\bibnamefont {Coleman}},\ }\href@noop {} {\emph {\bibinfo {title} {Introduction to Many-Body Physics}}}\ (\bibinfo  {publisher} {Cambridge University Press},\ \bibinfo {year} {2015})\BibitemShut {NoStop}%
\bibitem [{\citenamefont {Savary}\ and\ \citenamefont {Balents}(2017)}]{PhysRevLett.118.087203}%
  \BibitemOpen
  \bibfield  {author} {\bibinfo {author} {\bibfnamefont {L.}~\bibnamefont {Savary}}\ and\ \bibinfo {author} {\bibfnamefont {L.}~\bibnamefont {Balents}},\ }\href {\doibase 10.1103/PhysRevLett.118.087203} {\bibfield  {journal} {\bibinfo  {journal} {Phys. Rev. Lett.}\ }\textbf {\bibinfo {volume} {118}},\ \bibinfo {pages} {087203} (\bibinfo {year} {2017})}\BibitemShut {NoStop}%
\bibitem [{\citenamefont {Kwasigroch}\ \emph {et~al.}(2017)\citenamefont {Kwasigroch}, \citenamefont {Dou\ifmmode~\mbox{\c{c}}\else \c{c}\fi{}ot},\ and\ \citenamefont {Castelnovo}}]{PhysRevB.95.134439}%
  \BibitemOpen
  \bibfield  {author} {\bibinfo {author} {\bibfnamefont {M.~P.}\ \bibnamefont {Kwasigroch}}, \bibinfo {author} {\bibfnamefont {B.}~\bibnamefont {Dou\ifmmode~\mbox{\c{c}}\else \c{c}\fi{}ot}}, \ and\ \bibinfo {author} {\bibfnamefont {C.}~\bibnamefont {Castelnovo}},\ }\href {\doibase 10.1103/PhysRevB.95.134439} {\bibfield  {journal} {\bibinfo  {journal} {Phys. Rev. B}\ }\textbf {\bibinfo {volume} {95}},\ \bibinfo {pages} {134439} (\bibinfo {year} {2017})}\BibitemShut {NoStop}%
\bibitem [{\citenamefont {Coleman}\ and\ \citenamefont {Andrei}(1986)}]{PColeman_1986}%
  \BibitemOpen
  \bibfield  {author} {\bibinfo {author} {\bibfnamefont {P.}~\bibnamefont {Coleman}}\ and\ \bibinfo {author} {\bibfnamefont {N.}~\bibnamefont {Andrei}},\ }\href {\doibase 10.1088/0022-3719/19/17/017} {\bibfield  {journal} {\bibinfo  {journal} {Journal of Physics C: Solid State Physics}\ }\textbf {\bibinfo {volume} {19}},\ \bibinfo {pages} {3211} (\bibinfo {year} {1986})}\BibitemShut {NoStop}%
\bibitem [{\citenamefont {Coleman}\ and\ \citenamefont {P\'epin}(2003)}]{PhysRevB.68.220405}%
  \BibitemOpen
  \bibfield  {author} {\bibinfo {author} {\bibfnamefont {P.}~\bibnamefont {Coleman}}\ and\ \bibinfo {author} {\bibfnamefont {C.}~\bibnamefont {P\'epin}},\ }\href {\doibase 10.1103/PhysRevB.68.220405} {\bibfield  {journal} {\bibinfo  {journal} {Phys. Rev. B}\ }\textbf {\bibinfo {volume} {68}},\ \bibinfo {pages} {220405} (\bibinfo {year} {2003})}\BibitemShut {NoStop}%
\bibitem [{\citenamefont {Anderson}(1967)}]{PhysRevLett.18.1049}%
  \BibitemOpen
  \bibfield  {author} {\bibinfo {author} {\bibfnamefont {P.~W.}\ \bibnamefont {Anderson}},\ }\href {\doibase 10.1103/PhysRevLett.18.1049} {\bibfield  {journal} {\bibinfo  {journal} {Phys. Rev. Lett.}\ }\textbf {\bibinfo {volume} {18}},\ \bibinfo {pages} {1049} (\bibinfo {year} {1967})}\BibitemShut {NoStop}%
\bibitem [{\citenamefont {Nozi\`eres}\ and\ \citenamefont {De~Dominicis}(1969)}]{PhysRev.178.1097}%
  \BibitemOpen
  \bibfield  {author} {\bibinfo {author} {\bibfnamefont {P.}~\bibnamefont {Nozi\`eres}}\ and\ \bibinfo {author} {\bibfnamefont {C.~T.}\ \bibnamefont {De~Dominicis}},\ }\href {\doibase 10.1103/PhysRev.178.1097} {\bibfield  {journal} {\bibinfo  {journal} {Phys. Rev.}\ }\textbf {\bibinfo {volume} {178}},\ \bibinfo {pages} {1097} (\bibinfo {year} {1969})}\BibitemShut {NoStop}%
\bibitem [{\citenamefont {Peters}\ \emph {et~al.}(2012)\citenamefont {Peters}, \citenamefont {Kawakami},\ and\ \citenamefont {Pruschke}}]{PhysRevLett.108.086402}%
  \BibitemOpen
  \bibfield  {author} {\bibinfo {author} {\bibfnamefont {R.}~\bibnamefont {Peters}}, \bibinfo {author} {\bibfnamefont {N.}~\bibnamefont {Kawakami}}, \ and\ \bibinfo {author} {\bibfnamefont {T.}~\bibnamefont {Pruschke}},\ }\href {\doibase 10.1103/PhysRevLett.108.086402} {\bibfield  {journal} {\bibinfo  {journal} {Phys. Rev. Lett.}\ }\textbf {\bibinfo {volume} {108}},\ \bibinfo {pages} {086402} (\bibinfo {year} {2012})}\BibitemShut {NoStop}%
\bibitem [{\citenamefont {\"Ostlund}(2007)}]{PhysRevB.76.153101}%
  \BibitemOpen
  \bibfield  {author} {\bibinfo {author} {\bibfnamefont {S.}~\bibnamefont {\"Ostlund}},\ }\href {\doibase 10.1103/PhysRevB.76.153101} {\bibfield  {journal} {\bibinfo  {journal} {Phys. Rev. B}\ }\textbf {\bibinfo {volume} {76}},\ \bibinfo {pages} {153101} (\bibinfo {year} {2007})}\BibitemShut {NoStop}%
\end{thebibliography}%
\appendix

\section{Large-$N$ Newns-Read theory of underscreened $S=1$ Kondo Lattice}

\label{app:gauge}

In this section, we generalise the large-$N$ Read-Newns mean-field theory to the underscreened $S=1$ Kondo lattice. We examine the fixed-gauge $N\rightarrow\infty$ saddle point solution and the fluctuations around it. 
The partition function is given by
\begin{figure*}
    \centering
    \includegraphics[width=0.95\textwidth]{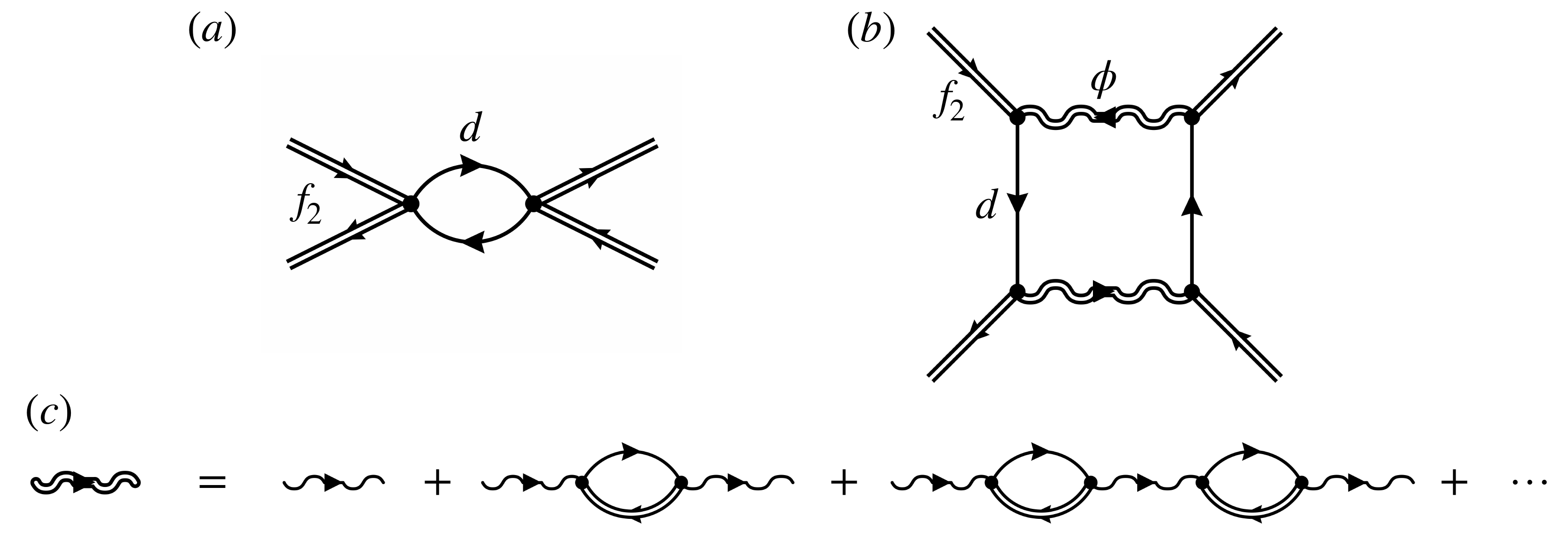}
    \caption{Feynman diagrams for the (a) RKKY interaction between unhybridized $f_2$-fermions, (b) the Hund's interaction between $f_2$-fermions, and (c) the diagrammatic expansion of the $\phi$ propagator $G_{\phi\phi}(\omega)=\langle \phi^\dagger(\omega)\phi(\omega)\rangle$ in the RPA approximation. The double lines indicate the unhybridised $f_2$-fermions, $d$ are the heavy fermions (superposition of the hybridized $f_1$-fermions and conduction electrons) and wavy lines are the $\phi$ gauge fields.}
    \label{fig:Feynman}
\end{figure*}
\begin{widetext}
\begin{align}
    \mathcal{Z} =&  \left(\prod_{a \alpha\sigma i}\int \mathcal{D} f_{a\alpha\sigma}^{\dagger}(\tau,\mathbf{r}_i) \;
    \mathcal{D} f_{a\alpha\sigma}(\tau,\mathbf{r}_i)\;
    \mathcal{D} c_{\alpha\sigma}^{\dagger}(\tau,\mathbf{r}_i)\;
    \mathcal{D} c_{\alpha\sigma}(\tau,\mathbf{r}_i)\right)\;
   \left( \prod_{a i} 
   \int \mathcal{D} V_{a i }^{\dagger}(\tau)
    \mathcal{D} V_{a i }(\tau)
   \right)\times\\
    &\left(
    \prod_{i\eta} 
    \int \mathcal{D} \phi^{\eta}_i(\tau)
    \right) \left(
    \prod_{i} 
    \int \mathcal{D} \theta_i(\tau)
    \right)
    e^{-\mathcal{S}}, 
    \nonumber\\
\mathcal{S} &= \int_{0}^{\beta} \; {\rm d}\tau  \sum_{\mathbf{k} a\alpha\sigma}
    \left(f_{a\alpha\sigma}^{\dagger}(\mathbf{k}) \partial_{\tau} f_{a\alpha\sigma}(\mathbf{k})+c_{\alpha\sigma}^{\dagger}(\mathbf{k}) \partial_{\tau} c_{\alpha \sigma}(\mathbf{k}) + \epsilon_{\mathbf{k}} c_{\alpha\sigma}^{\dagger}(\mathbf{k}) c_{\alpha\sigma}(\mathbf{k})
    \right)
    \nonumber\\
    &+\int_{0}^{\beta} \; {\rm d}\tau  \sum_{i, a}
    \left(
    J_{ K}N V_{a i}^{\dagger} V_{a i}
    +J_{ K} \left( V_{a  i} \sum_{\alpha\sigma}
    f_{a\alpha\sigma}^{\dagger}(\mathbf{r}_i)
    c_{\alpha\sigma}(\mathbf{r}_i) +{\rm h.c.}
    \right)\right)
    \nonumber\\
    &\;\;+
    \int_{0}^{\beta} \; {\rm d}\tau  \sum_{i}
    i\theta_i \left(
    \sum_{a\alpha} f_{a \alpha\sigma}^{\dagger}(\mathbf{r}_i)f_{a \alpha\sigma}(\mathbf{r}_i)
   -N \right)+
   \int_{0}^{\beta} \; {\rm d}\tau \sum_{i\eta} i\phi^{\eta}_i
  \sum_{a a'\alpha\sigma} f^{\dagger}_{a\alpha\sigma}(\mathbf{r}_i) \tau^{\eta}_{a a'} f_{a'\alpha\sigma}(\mathbf{r}_i).
    \nonumber
\end{align}
\end{widetext}
The general hybridisation field on site $i$ and time $\tau$ can be written in terms of radial and angular coordinates $|V_i|,\Theta_i,\Phi^x_i,\Phi^y_i,\Phi_i^z$. Since this introduces too many coordinates, we make the choice $\Phi_i^z(\tau)=\Theta_i(\tau)$.
Through a series of $U(2)$ transformations given by the matrix $\mathcal{U}$, (all variables in general depend on $\tau$) we can decouple the $f_2$-moments from the hybridisation field as follows
\begin{widetext}
\begin{align}
    \begin{pmatrix}
        V_{1i} \\ V_{2i}
    \end{pmatrix}
    &=
    \begin{pmatrix}
   \cos\Phi^x_i\cos\Phi^y_i + i\sin\Phi^x_i\sin\Phi^y_i
    \\
    \cos\Phi^x\sin\Phi^y_i+i\sin\Phi^x_i\cos\Phi^y_i
    \end{pmatrix}
    |V_i|e^{i\Theta_i+i \Phi^z_i}
    \nonumber\\
    &=
    e^{i \Theta_i} 
     \begin{pmatrix}
    \cos \Phi^x_i && i\sin \Phi^x_i\\
    i\sin\Phi^x_i && \cos\Phi^x_i
    \end{pmatrix}
    \begin{pmatrix}
    \cos \Phi^y_i && -\sin \Phi^y_i\\
    \sin\Phi^y_i && \cos\Phi^y_i
    \end{pmatrix}
    \begin{pmatrix}
         e^{i \Phi^z_i} && 0
         \\
         0 && e^{-i \Phi^z_i}
    \end{pmatrix}
     \begin{pmatrix}
        |V_i| \\ 0
    \end{pmatrix} : = 
    \mathcal{U}\begin{pmatrix}
        |V_i| \\ 0
    \end{pmatrix}\nonumber\\
    \begin{pmatrix}
        f_{1\alpha\sigma}(\mathbf{r}_i) \\ f_{2\alpha\sigma}(\mathbf{r}_i)
    \end{pmatrix} &\rightarrow \mathcal{U}
   \begin{pmatrix}
        f_{1\alpha\sigma}(\mathbf{r}_i) \\ f_{2\alpha\sigma}(\mathbf{r}_i)
    \end{pmatrix},
    \nonumber\\
    \begin{pmatrix}
        \phi^x_i \\ \phi^y_i \\ \phi^z_i
    \end{pmatrix} &\rightarrow 
     \begin{pmatrix}
        \cos2\Phi^z_i & -\sin 2\Phi^z_i & 0
        \\
        \sin2\Phi^z_i & \cos 2\Phi^z_i & 0
        \\
        0 & 0 & 1
    \end{pmatrix}
    \begin{pmatrix}
        \cos2\Phi^y_i & 0 & -\sin 2\Phi^y_i 
        \\
        0  & 1 & 0
        \\
        \sin2\Phi^y_i & 0 & \cos 2\Phi^y_i 
        \\
    \end{pmatrix}
    \begin{pmatrix}
       1  & 0 & 0
       \\
        0 &\cos2\Phi^x_i & \sin 2\Phi^x_i     
        \\
        0 &-\sin2\Phi^x_i &  \cos 2\Phi^x_i 
        \\
    \end{pmatrix}
   \begin{pmatrix}
       \phi^x_i \\ \phi^y_i \\ \phi^z_i
    \end{pmatrix}, \label{eq:transformation}
\end{align}
followed by
\begin{eqnarray}
    \theta_i &\rightarrow& \theta_i - \partial_{\tau}\Theta_i,
    \nonumber\\
  \phi^z_i &\rightarrow& \phi^z_i- \;\partial_{\tau}\Phi^z_i-i\sin2\Phi^y_i\partial_\tau \Phi^x_i,
  \nonumber\\
  \phi^x_i &\rightarrow& \phi^x_i- \cos2\Phi^y_i \cos2\Phi^z_i \;\partial_{\tau}\Phi^x_i-\sin2\Phi^z_i\partial_\tau\Phi^y_i,
  \nonumber\\
  \phi^y_i&\rightarrow&\phi^y_i +\cos2\Phi_i^z \partial_\tau \Phi^y_i-\partial_{\tau}\Phi^x_i\cos2\Phi^y_i\sin2\Phi^z_i. \label{eq: dynamic fluctuations}
\end{eqnarray}
The action becomes 
\begin{align}
\mathcal{S} &= \int_{0}^{\beta} \; {\rm d}\tau  \sum_{\mathbf{k} a\alpha\sigma}
    \left(f_{a\alpha\sigma}^{\dagger}(\mathbf{k}) \partial_{\tau} f_{a\alpha \sigma}(\mathbf{k})+c_{\alpha\sigma}^{\dagger}(\mathbf{k}) \partial_{\tau} c_{\alpha \sigma}(\mathbf{k}) + \epsilon_{\mathbf{k}} c_{\alpha\sigma}^{\dagger}(\mathbf{k}) c_{a\alpha\sigma}(\mathbf{k})
    \right)
    \nonumber\\
    &+\int_{0}^{\beta} \; {\rm d}\tau  \sum_{i}
    \left(
    J_{ K}N |V_i|^2
    +J_{ K} \left( |V_{  i}| \sum_{\alpha\sigma}
    f_{1\alpha\sigma}^{\dagger}(\mathbf{r}_i)
    c_{\alpha\sigma}(\mathbf{r}_i) +{\rm h.c.}
    \right)\right)
    \nonumber\\
    &\;\;+
    \int_{0}^{\beta} \; {\rm d}\tau  \sum_{i}
    i\theta_i \left(
    \sum_{a\alpha\sigma} f_{a \alpha\sigma}^{\dagger}(\mathbf{r}_i)f_{a \alpha\sigma}(\mathbf{r}_i)
   -N \right)+
   \int_{0}^{\beta} \; {\rm d}\tau \sum_{i} i\phi^{\eta}_i
  \sum_{a a'\alpha\sigma} f^{\dagger}_{a\alpha\sigma}(\mathbf{r}_i) \tau^{\eta}_{a a'} f_{a'\alpha\sigma}(\mathbf{r}_i).
    \nonumber\\
\end{align}
\end{widetext}
The action now depends only on the magnitude of the hybridisation field $|V_i|$ and $f_2$-moments are no longer directly coupled to it. Since the action no longer depends on the angular coordinates describing the hybridisation field in Eq. \ref{eq:transformation}, we can set $\Phi^y_i=\Phi^z_i=\Theta_i=0$ and write the partition function as an integral over $|V_i|$ only
\begin{widetext}
\begin{align}
    \mathcal{Z} =  &\left(\prod_{a \alpha\sigma i}\int \mathcal{D} f_{a\alpha\sigma}^{\dagger}(\tau,\mathbf{r}_i) \;
    \mathcal{D} f_{a\alpha\sigma}(\tau,\mathbf{r}_i)\;
    \mathcal{D} c_{\alpha\sigma}^{\dagger}(\tau,\mathbf{r}_i)\;
    \mathcal{D} c_{\alpha\sigma}(\tau,\mathbf{r}_i)\right)\;
    \times
    \nonumber\\
   &\left( \prod_{i} 
   \int |V_{i }(\tau)|^3\mathcal{D} |V_{i }(\tau)|
   \right)
    \left(
    \prod_{i\eta} 
    \int \mathcal{D} \phi^{\eta}_i(\tau)
    \right) \left(
    \prod_{i} 
    \int \mathcal{D} \theta_i(\tau)
    \right)
    e^{-\mathcal{S}}.
    \nonumber
\end{align}
\end{widetext}
It is convenient to re-express the gauge fields in terms of new variables
\begin{eqnarray}
    \lambda_{1i} &=& i \theta_i + i\phi^z_i,
    \nonumber\quad
    \lambda_{2i} = i \theta_i - i\phi^z_i,
     \nonumber\\
    \phi &=& \phi^x - i\phi^y_i,
    \quad
    \phi^{\dagger} = \phi^x + i\phi^y_i,
\end{eqnarray}
so that the action and partition function become (up to a constant multiplicative prefactor)
\begin{widetext}
\begin{eqnarray}
    \mathcal{Z} &=&  \left(\prod_{a \alpha\sigma i}\int \mathcal{D} f_{a\alpha\sigma }^{\dagger}(\tau,\mathbf{r}_i) \;
    \mathcal{D} f_{a\alpha\sigma}(\tau,\mathbf{r}_i)\;
    \mathcal{D} c_{\alpha\sigma}^{\dagger}(\tau,\mathbf{r}_i)\;
    \mathcal{D} c_{\alpha\sigma}(\tau,\mathbf{r}_i)\right)\;
   \left( \prod_{i} 
   \int \mathcal{D} |V_{i }(\tau)|
   \right)
    \left(
    \prod_{i} 
    \int \mathcal{D} \phi^{\dagger}_i(\tau) \mathcal{D} \phi^{}_i(\tau)
    \right)
    \nonumber\\
   && \times
    \left(
    \prod_{i} 
    \int \mathcal{D} \lambda_{1i}(\tau)
    \right)\left(
    \prod_{i} 
    \int_0^{2\pi i T} {\rm d} \lambda_{2i}
    \right)
    e^{-\mathcal{S}}, 
    \nonumber
\end{eqnarray}
\begin{eqnarray}
\mathcal{S} &=& \int_{0}^{\beta} \; {\rm d}\tau  \sum_{\mathbf{k} a \alpha\sigma}
    \left(f_{a\alpha\sigma}^{\dagger}(\mathbf{k}) \partial_{\tau} f_{a\alpha\sigma}(\mathbf{k}) +c_{\alpha\sigma}^{\dagger}(\mathbf{k})  \partial_{\tau} c_{\alpha\sigma}(\mathbf{k})  + \epsilon_{\mathbf{k}} c_{\alpha\sigma}^{\dagger}(\mathbf{k})  c_{\alpha \sigma}(\mathbf{k}) 
    \right)
    \nonumber\\
    &&+\int_{0}^{\beta} \; {\rm d}\tau  \sum_{i}
    \left(
    J_{ K}N |V_i|^2
    +J_{ K} \left( |V_{  i}| \sum_{\alpha\sigma}
    f_{1\alpha\sigma}^{\dagger}(\mathbf{r}_i)
    c_{\alpha\sigma}(\mathbf{r}_i) +{\rm h.c.}
    \right)\right)
    \nonumber\\
    &&\;\;+
    \int_{0}^{\beta} \; {\rm d}\tau  \sum_{ia}
    \lambda_{a i} \left(
    \sum_{\alpha\sigma} f_{a \alpha\sigma}^{\dagger}(\mathbf{r}_i)f_{a \alpha\sigma}(\mathbf{r}_i)
   -\frac{N}{2} \right)+
   \int_{0}^{\beta} \; {\rm d}\tau \sum_{i \alpha\sigma}
  i \left(
   \phi^{}_i f^{\dagger}_{1\alpha\sigma}(\mathbf{r}_i) f_{2\alpha\sigma}(\mathbf{r}_i) + {\rm h.c.}\right),
    \nonumber\\
\end{eqnarray}
\end{widetext}
where we have only kept the static and bounded part of $\lambda_{2i}(\tau)$ (see later for details).

We now consider the fluctuations around the fixed-gauge $N\rightarrow\infty$ saddle-point
\begin{eqnarray}
    \lambda_{1 i}(\tau)&=& \bar{\lambda}_1+ i\delta \lambda_{1 i}(\tau),
    \nonumber\\
      |V_{i}(\tau)|&=& \bar{V}_1+ \delta V_{ 1i}(\tau),
      \nonumber\\
        \phi_{ i}(\tau)&=& 0+ \delta \phi_{ i}(\tau),
\end{eqnarray}
where $\bar{\lambda}_1$ and $\bar{V}_1$ are chosen to minimise the free energy.  In the Gaussian approximation, working to quadratic order in the fluctuations (i.e. to $\mathcal{O}\left(\frac{1}{N}\right)$), we integrate out the fermionic fields and obtain the fluctuation correction to the $N\rightarrow\infty$ saddle-point action 
\begin{widetext}
\begin{align}
\nonumber
    \delta\mathcal{S} = \sum_{\omega, i} \left(\frac{1}{2}G^{-1}_{VV}(\omega)\delta V_{ 1 i}(\omega)\delta V_{1 i} (-\omega)+ G^{-1}_{V\lambda}(\omega)\delta V_{1 i}(\omega) \delta \lambda_{1 i}(-\omega)+
    \frac{1}{2}G^{-1}_{\lambda\lambda}(\omega)\delta \lambda_{ 1 i}(\omega)\delta \lambda_{ 1 i}(-\omega)+
    G^{-1}_{\phi \phi}(\omega)\delta \phi^{\dagger}_{  i}(\omega)\delta \phi_{  i}(\omega) \right),
\end{align}
where we have focused on the frequency dependence and have therefore omitted terms coupling different lattice sites.
Except for $ G^{-1}_{\phi\phi}(0)$, the low-energy limits ($\omega\rightarrow 0$) of the propagators are the same as those for the standard $S=1/2$ Kondo lattice large-$N$ theory and have been examined previously \cite{PhysRevLett.57.877}. Here, we present the result for the propagator of the bosonic gauge field associated with the Hund constraint coupling $f_1$- and $f_2$-fermions
\begin{eqnarray}
    G^{-1}_{\phi\phi}(0) &=& \frac{1}{\beta}
    \sum_{\alpha\sigma\sigma'}
\int_0^{\beta} \int_0^{\beta} 
\langle f^{\dagger}_{1 \alpha\sigma}(\tau,\mathbf{r}_i)  f_{2 \alpha\sigma}(\tau,\mathbf{r}_i)
f^{\dagger}_{2 \alpha\sigma'}(\tau',\mathbf{r}_i)  f_{1 \alpha\sigma'}(\tau',\mathbf{r}_i)
 \rangle_0 \;
 d \tau d \tau' 
 \nonumber\\
 &=&
\frac{1}{2\beta N_s} \sum_{\alpha\sigma \mathbf{k}}
\left(  
\int_0^{\beta} \int_{\tau'}^{\beta} 
\langle
f^{\dagger}_{1 \alpha\sigma}(\tau,\mathbf{k})   f_{1 \alpha \sigma }(\tau',\mathbf{k})\rangle_0 \;d\tau \; d\tau'
+ \int_0^{\beta} \int_{\tau}^{\beta} 
\langle
 f_{1 \alpha\sigma}(\tau',\mathbf{k}) f^{\dagger}_{1 \alpha \sigma}(\tau,\mathbf{k})  
 \rangle_0 \; d\tau' \; d\tau
 \right)
\nonumber\\
&=&
\frac{N\rho}{2}\int_{\rm bands} \left(\frac{J_K\bar{V}_1}{E-\bar{\lambda}_1}\right)^2 \frac{n(-E)-n(E)}{E}  d E
\sim N\rho \left(\frac{J_K\bar{V}_1}{\bar{\lambda}_1}\right)^2\log \frac{\bar{\lambda}_1}{T},
\end{eqnarray}
\end{widetext}
where the final integral is over the $N\rightarrow\infty$ saddle-point energy bands calculated in App. \ref{UVA}. Away from the insulating limit, we have isolated the divergent contribution to the propagator, arising from integration around the Fermi level.  The low-energy fluctuations in the $\phi_i(\tau)$ field now generate a residual ferromagnetic interaction between the unhybridised $f_2$-moments and heavy-fermions ($f_1$-moments can now be expressed in terms of the heavy-fermion quasiparticles)
\begin{eqnarray}\label{fluc}
    &\Delta H_{\rm fluc}  \sim 
    \nonumber\\&- G_{\phi\phi}(0)
    \sum_{\alpha \beta\sigma\sigma' i  }f_{1\alpha\sigma}^{\dagger}(\mathbf{r}_i) f_{1\beta\sigma'}(\mathbf{r}_i)  f_{2\beta\sigma'}^{\dagger}(\mathbf{r}_i)  f_{2\alpha\sigma}(\mathbf{r}_i).\label{eq:Hund}
    \nonumber\\
\end{eqnarray}
We have used the RPA approximation to calculate the propagator
\begin{align}
    \langle \phi_i^{\dagger}(\omega) \phi_i(\omega)\rangle
    =
    \frac{G^{0}_{\phi\phi}(\omega)}{1-(i)^2G^{-1}_{\phi\phi}(\omega)G^0_{\phi\phi}(\omega)}\rightarrow G_{\phi\phi}(\omega),
\end{align}
since the bare propagator $G^{0}_{\phi\phi}(\omega)\rightarrow\infty$ (the stiffness of the $\phi_i(\tau)$ is wholly generated by virtual hopping between $a=1,2$ fermionic states).
The propagator in the RPA approximation is depicted in Fig. \ref{fig:Feynman}(c), and its $\omega=0$ limit gives the strength of the residual logarithmic interaction in $\Delta H_{\rm fluc}$.
Note that such a logarithmic residual interaction has been found to give rise to singular Fermi-liquid behaviour (signalled by, e.g., a logarithmically divergent heat capacity coefficient) \cite{PhysRevB.68.220405}. However, once the impurity becomes polarised (e.g. in the presence of an external magnetic field or magnetic order) conventional Fermi-liquid behaviour is restored. The logarithmic exchange between the residual moment and heavy-fermions also gives rise to an exchange, which we have dubbed 'Hund' exchange, between $f_2$-fermions on different lattice sites, once itinerant quasiparticles are integrated out. The exchange is depicted in Fig. \ref{fig:Feynman}(b) and calculated in App. \ref{sec:limit}. It is however outcompeted by RKKY exchange, which we discuss next, and becomes irrelevant at low energies.

To linear order in the fields, the dynamic fluctuations of the gauge fields $\theta_i(\tau)$ and $\phi^{\eta}_i(\tau)$ correspond to the transverse fluctuations of the hybridization field, i.e., they correspond to the time variation of the angle variables $\Theta_i(\tau)$ and $\Phi^{\eta}_i(\tau)$, respectively, as shown by Eq. \ref{eq: dynamic fluctuations}. Although, for simplicity, we have used fully time-dependent gauge field variables to begin with in our partition function, we could have started with static (time-independent) and bounded variables $\theta_i,|\boldsymbol{\phi}_i| \leq 2\pi T$, which are introduced into the partition function via the Gutzwiller projection operator defined in Eq. \ref{eq:Gutzwiller}. Except for the $\lambda_{2i}$ gauge field, as $T\rightarrow 0$, this static and bounded component becomes irrelevant and the gauge fields follow the dynamics of the angle variables, that describe the transverse fluctuations of the hybridization field, via the transformation defined in Eqs. \ref{eq:transformation} - \ref{eq: dynamic fluctuations}. 
From Eq. \ref{eq: dynamic fluctuations}, we can see that our choice of $\Phi^z_i(\tau)=\Theta_i(\tau)$ leads to a cancellation of time-dependent fluctuations so that $\lambda_{2i}=i\theta_i(\tau)-i\phi^z(\tau)$ is static and bounded to first order in the fields. Integration over the static $\lambda_{2i} \in [0,2\pi)$ gauge field is equivalent to the action of a Gutzwiller operator that projects out states which violate $\hat{n}^f_{2i}=1$ from the partition function.
The non-zero stiffness of the remaining $\lambda_{1i}(\tau)$ and $\phi_i(\tau)$ gauge fields then translates to a finite decoherence of the hybridization field via long-wavelength transverse fluctuations. Starting with the hybridization field aligned along the $a=1$ direction and purely real at $\tau=0$, the power-law decoherence can be derived as follows
\begin{widetext}
\begin{align}
    \sum_a\langle V_{ai}^{\dagger}(\tau)V_{ai}(0)\rangle
    &=\bar{V}_{1i}^2\langle e^{i\Phi^y_i(\tau) -\Phi^y_i(0)}\rangle 
    \langle e^{i\Phi^x_i(\tau) -\Phi^x_i(0)}\rangle 
    \langle e^{(i\Theta_i(\tau)+i\Phi_i^z(\tau)-i\Theta_i(0)-i\Phi^z_i(0))}\rangle
    \nonumber\\
&=\bar{V}_{1i}^2
e^{\langle\Phi_i^x(\tau)\Phi_i^x(0)-\Phi^x_i(0)^2\rangle }
e^{\langle\Phi_i^y(\tau)\Phi_i^y(0)-\Phi^y_i(0)^2\rangle }
    e^{\langle(\Theta_i(\tau)+\Phi^z_i(\tau))(\Theta_i(0)+\Phi^z_i(0))-(\Theta_i(0)+\Phi^z_i(0))^2\rangle}
    \nonumber\\
     &=\bar{V}_{1i}^2
     e^{\int \frac{G_{\phi\phi}(\omega)+G_{\lambda\lambda}(\omega)}{\omega^2} (\cos\omega\tau-1){\rm d}\omega } \sim\frac{1}{\tau^{1/N}},
\end{align}
\end{widetext}
where we have neglected the longitudinal fluctuations of the hybridization field and the contribution from the fluctuations of the infinitely stiff $\phi_i(\tau)$ gauge field ($G_{\phi\phi}(\omega\rightarrow 0)=0$), and computed the expectation values in the Gaussian approximation. The infinitely stiff $\phi_i(\tau)$ gauge field quenches long-wavelength (in time), transverse fluctuations of the hybridization field along the $a=2$ direction, i.e.,  $V_{2i}(\tau)$. This leads to the logarithmic decoupling of $f_2$-fermions in the hybridization channel, which relies on long-wavelength coherence. However, the exchange-channel interaction between $f_2$-fermions and conduction electrons is generated by short wavelength (incoherent in time) fluctuations of  $V_{2i}(\tau)$,  whose energy cost in the action is simply given by $\int_{0}^{\beta}  \sum_{i}
    J_{ K}N V_{2i}^{\dagger}V_{2i} {\rm d}\tau =\int_{0}^{\beta}  \sum_{i}
    J_{ K}N \bar{V_{1}}^2 \Phi^{\dagger}\Phi \;{\rm d}\tau $, where one can show that $\omega^2G^{-1}_{\phi\phi}(\omega)\rightarrow J_KN \bar{V}_{1}^2$ for $\omega\gg T_K$. This gives rise to the following fluctuation correction to the Hamiltonian
    \begin{eqnarray}\label{fluc}
    &\Delta H_{\rm fluc}  = 
    \nonumber\\&
    \frac{J_K}{N}\sum_{\alpha \beta\sigma\sigma' i  }c_{\alpha\sigma}^{\dagger}(\mathbf{r}_i) c_{\beta\sigma'}(\mathbf{r}_i)  f_{2\beta\sigma'}^{\dagger}(\mathbf{r}_i)  f_{2\alpha\sigma}(\mathbf{r}_i),
    \nonumber\\
\end{eqnarray}
with the important caveat that long-wavelength (in imaginary time), Kondo-channel, processes are excluded. This leaves a pure exchange-channel interaction that can be treated perturbatively.
    Once itinerant fermions are integrated out to second order in $1/N$, we obtain an RKKY interaction between the unhybridized $f_2$-fermions, shown in the Feynman diagram in Fig. \ref{fig:Feynman}(a), and given in App. \ref{sec:limit}.
\section{Mean field theory}\label{UVA}
The mean field order parameters can be expressed in the terms of the matrix elements which diagonalise the Hamiltonian as follows:
\begin{eqnarray}
    \bar{V}_1 &=& \frac{1}{N N_s} \sum_{\alpha\sigma\mathbf{k} } \langle f^{\dagger}_{1\alpha\sigma}(\mathbf{k}) c_{\alpha\sigma}(\mathbf{k})\rangle
    = \frac{1}{N N_s} \sum_{\sigma\mathbf{k} } v^{-}_{\mathbf{k}\sigma} u^{-}_{\mathbf{k}\sigma},
    \nonumber\\
\bar{S}_1 &=& \frac{2}{N N_s}\sum_{\alpha\sigma \mathbf{k}} \sigma \langle
f^{\dagger}_{1\alpha\sigma}(\mathbf{k}) 
f_{1\alpha\sigma}(\mathbf{k}) 
\rangle = \frac{2}{N N_s}\
\sum_{\sigma\mathbf{k}} \sigma \left( v^{-}_{\mathbf{k} \sigma} \right)^2,
    \nonumber\\
\bar{s} &=& \frac{2}{N N_s}\sum_{\alpha\sigma \mathbf{k}} \sigma \langle
c^{\dagger}_{\alpha\sigma}(\mathbf{k}) 
c_{\alpha\sigma}(\mathbf{k}) 
\rangle\! =\!\! \frac{2}{N N_s}\
\!\!\sum_{\sigma \mathbf{k}} \sigma \left(u^{-}_{\mathbf{k}\sigma}\right)^2,
\nonumber\\
\end{eqnarray}
where we have assumed a partially filled ($\nu=-$) lower band and $\bar{S}_2=\frac{1}{2}$ in the ground state. Motivated by the large-$N$ expansion, where the $\lambda_{1i}$ gauge field develops a non-zero stiffness, allowing for non-zero fluctuations of the $\hat{n}^f_{1i}$ operator, we will soften the constraint on the number of $f_1$-fermions per site being $N/2$ and demand that it is only satisfied on average.  We thus want to minimise the energy subject to the constraint that average number of $f_1$-electrons per site is $\bar{n}_1=\frac{N}{2}$. We also want the average number of conduction electrons per site to be $\bar{n}_c=\frac{Nn_c}{2}$, where $n_c$ describes the conduction electron filling. The relevant expectation values are given below
\begin{eqnarray}
    \bar{n}_1 &=& \frac{1}{N_s}\sum_{\mathbf{k}\alpha\sigma} 
    \langle f^{\dagger}_{1\alpha\sigma}(\mathbf{k}) f_{1\alpha\sigma}(\mathbf{k})\rangle
    =\frac{1}{ N_s} \sum_{\mathbf{k}\sigma} \left(v^{-}_{\mathbf{k}\sigma}\right)^2,
    \nonumber\\
    \bar{n}_c &=& \frac{1}{N_s}\sum_{\mathbf{k}\alpha\sigma} 
    \langle c^{\dagger}_{\alpha\sigma}(\mathbf{k}) c_{\alpha\sigma}(\mathbf{k})\rangle
    =\frac{1}{ N_s} \sum_{\mathbf{k}\sigma} \left(u^{-}_{\mathbf{k}\sigma}\right)^2.
\end{eqnarray}
To fix the constraints on the occupancy of the conduction and impurity fermions we introduce the Lagrange multipliers $\bar{\lambda}_1, \mu$ which can be interpreted as the chemical potentials for the $f_1$-fermions (up to a minus sign) and conduction electrons respectively. We obtain
\begin{eqnarray}\label{transform}
    u_{\mathbf{k} \sigma}^{\pm} &=& 
    \frac{E^{\pm}_{\mathbf{k}\alpha} -\bar{\lambda}_{1\sigma}}{\sqrt{(E^{\pm}_{\mathbf{k}\sigma} -\bar{\lambda}_{1\sigma})^2 + J_K^2 \bar{V}_1^2}},
    \nonumber\\
       v_{\mathbf{k} \sigma}^{\pm} &=& 
    \frac{-J_K\bar{V}_1}{\sqrt{(E^{\pm}_{\mathbf{k}\sigma} -\bar{\lambda}_{1\sigma})^2 + J_K^2 \bar{V}_1^2}},
\end{eqnarray}
where 
\begin{eqnarray}
\bar{\lambda}_{1\sigma} &:=&
\bar{\lambda}_1 + \sigma J_M \bar{s},
\nonumber\\
    E^{\pm}_{\mathbf{k}\sigma} &=& \frac{1}{2} \left(
    (\epsilon_{\mathbf{k} } -\mu + J_M(\bar{S}_1+\bar{S}_2) \sigma+\bar\lambda_{1\sigma} 
    \right)
    \nonumber\\
    &&\!\!\!\!\!\!\!\!\!\!\!\!\!\!\!\!\!\!\!\!\!\!\!\!\!\!\!\!\!\!\pm \frac{1}{2} \sqrt{ \left(\epsilon_{\mathbf{k} } -\mu + J_M(\bar{S}_1+\bar{S}_2) \sigma - \bar{\lambda}_{1\sigma} 
    \right)^2 + 4J_K^2 \bar{V}_1^2}.
\end{eqnarray}
This is the transformation that diagonalises the matrix $\mathcal{H}$ given in the main text. There are two heavy-fermion bands:  $d^{-}_{\alpha\sigma}(\mathbf{k})$ is the fermionic annihilation operator for the lower band with energy dispersion $E^{-}_{\mathbf{k}\sigma}$, and $d^{+}_{\alpha\sigma}(\mathbf{k})$ is the annihilation operator for the upper band  with energy dispersion $E^{+}_{\mathbf{k}\sigma}$. Since we are describing a Kondo metal, we will work away from half-filling (Kondo insulator). We will work with a representative conduction electron filling of $n_c=0.8$, so that only the lower heavy-fermion bands ($E^-_{\mathbf{k}\sigma}$) are filled. We have checked that the results are qualitatively similar for other fillings in the non-insulating regime ($n_c\neq 1$).

To derive the self-consistency equations we have assumed a constant density of states for the conduction electrons of $\rho=\frac{1}{2\Lambda}$ with $|\epsilon_{\mathbf{k}}|\leq\Lambda$, and $E^-_{\sigma {\rm min}}$ is the bottom of the partially filled heavy-fermion bands, 
 i.e. the value of $E^-_{\mathbf{k}\sigma}$ when $\epsilon_{\mathbf{k}}=-\Lambda$. The density of states in the heavy-fermion bands is amplified by the hybridisation with the impurities and follows from the differential of the map between $E^{\pm}_{\mathbf{k}\sigma}$ and $\epsilon_{\mathbf{k}\sigma}$
\begin{eqnarray}
    \rho \; {\rm d}\epsilon_{\mathbf{k}\sigma} = \rho \left( 1+ \left(\frac{J_K\bar{V}_1}{E^{\pm}_{\mathbf{k}\sigma} -\bar{\lambda}_{1}}\right)^2 
    \right) {\rm d} E^{\pm}_{\mathbf{k}\sigma}.
\end{eqnarray}
Four out of the five parameters $\{\bar{V},\bar{S}_1,\bar{s}, \mu, \bar{\lambda}_1\}$ can be eliminated giving an implicit equation for $\bar{s}$ that is then solved numerically. The results are shown in the main paper. We also compared the energies of the unhybridised ($\bar{V}_1=0$) and hybridised ($\bar{V}_1\neq 0$) states to ascertain the precise location of any first-order transitions.

\section{Generalisation to $S>1$} \label{app:generalisation}
We want to generalise the large-$N$ analysis to impurity spins greater than $S=1$. We start by writing the spin-$S$ impurity as a sum of $2S$ spin-1/2 fermions
\begin{equation}
    \mathbf{S}_{\rm Total}=\sum_{a=1}^{2S}\mathbf{S}_a.
\end{equation}
To ensure that we have a faithful representation of the spin-$S$ algebra, we must consider the allowed states and the resulting constraints on the system. Writing the spin in the above way increases the size of the Hilbert space considerably and thus we only want the following $2S+1$ eigenstates of the $S^z$ operator which contribute to the physical Hilbert space:
\begin{widetext}
\begin{eqnarray}\nonumber
    |S\rangle&=&f^\dagger_{\uparrow 1}f^\dagger_{\uparrow 2} \dots f^\dagger_{\uparrow 2S}|0\rangle\\\nonumber
    |S-1\rangle &=&\sqrt{\frac{1}{2S}}\left( f^\dagger_{\downarrow 1}f^\dagger_{\uparrow 2}f^\dagger_{\uparrow 3}\dots f^\dagger_{\uparrow 2S}+f^\dagger_{\uparrow 1}f^\dagger_{\downarrow 2}f^\dagger_{\uparrow 3} \dots f^\dagger_{\uparrow 2S}+f^\dagger_{\uparrow 1}f^\dagger_{\uparrow 2}f^\dagger_{\downarrow 3} \dots f^\dagger_{\uparrow 2S}+\dots\right) |0\rangle\\\nonumber
    |S-2\rangle &=&\sqrt{\frac{1}{2S^2-S}}\left( f^\dagger_{\downarrow 1}f^\dagger_{\downarrow 2}
    f^\dagger_{\uparrow 3}\dots f^\dagger_{\uparrow 2S}+f^\dagger_{\uparrow 1}f^\dagger_{\downarrow 2}f^\dagger_{\downarrow 3}\dots f^\dagger_{\uparrow 2S}+\dots \right)|0\rangle\\ \nonumber
    &\vdots&\\
    |-S\rangle&=&f^\dagger_{\downarrow 1}f^\dagger_{\downarrow 2} \dots f^\dagger_{\downarrow 2S}|0\rangle
\end{eqnarray}
\end{widetext}
For each state in $\{|S-n\rangle\,:\,n\in\mathbb{N}_0,\,n\leq 2S\}$ the spin of $n$ fermions is flipped but, as the flipped spin could be chosen from any of the total of $2S$ fermions, there will be ${\rm C}(2S,n)$ contributions for each state, which need to be considered to properly normalise each state. To ensure we only look at the physical states we introduce the following constraints which screen out the unphysical sectors of the enlarged Hilbert space. After generating $N/2$ replicas for each fermion spin, there are two types of constraints: one generating $U(1)$ transformations
\begin{eqnarray}
    \sum_{\sigma \alpha a}f^\dagger_{\sigma\alpha a}f_{\sigma\alpha a}=2S\cdot\frac{N}{2},
\end{eqnarray}
 which constrains the total number of impurity fermions, and those generating $SU(2S)$ transformations
\begin{eqnarray}
    \underbrace{\sum_{\sigma\alpha} \left( f^\dagger_{\sigma\alpha a}f_{\sigma\alpha a}-f^\dagger_{\sigma\alpha, a+1}f_{\sigma\alpha, a+1}\right)=0,}_{2S-1{\rm \ Constraints}}\\
    \underbrace{\sum_{\sigma\alpha} \left( f^\dagger_{\sigma\alpha a}f_{\sigma\alpha b}\pm {\rm h.c.}\right)=0,}_{C(2S,2)\cdot 2=4S^2-2S{\rm \ Constraints\ for\ }a\neq b}
\end{eqnarray}
where $a,b\in[1,2S]$, which ensure that the number of fermions for each index $a$ is the same and there is perfect Hund's coupling between spins $\mathbf{S}_a$ with different index $a$. These constraints together generate the $U(1)\times SU(2S)\sim U(2S)$ gauge symmetry group which can be exploited to fix the gauge of the hybridisation field as in the $S=1$ case. Furthermore, it can be shown that if the number of conduction channels is $N_c$ then you can exploit the gauge symmetry to ensure $N_c$ of the impurity fermions are decoupled from the hybridisation field after gauge fixing, and the total residual magnetic moment that is free to magnetise would be $S-N_c/2$.

\section{ RKKY interaction and order parameters in the $N\rightarrow\infty$ limit} \label{sec:limit}
At sufficiently large $N$, we have a separation of scales, whereby $T_{\rm RKKY}\ll T_K$. We can then treat the magnetic RKKY channel contributions to the ground state energy perturbatively. In particular, the RKKY interaction along the $z$-direction, generated between the unhybridised $f_2$-moments, is given by
\begin{eqnarray}
  \frac{\langle H_{\rm RKKY}\rangle}{N}  &=& -\frac{J_K^2}{ N^2 N_S} \sum_{\mathbf{q}, i j}\chi_{\rm cc}(\mathbf{q})  e^{i \mathbf{q} \cdot(\mathbf{r}_i-\mathbf{r}_j)} \bar{S}_{2i}\bar{S}_{2j},
  \nonumber\\
\end{eqnarray}
where we have made the usual static approximation with $f_2$-moments treated as classical spins (we have assumed the same magnetization for each replica). The $N\rightarrow\infty$ susceptibilities are given by
\begin{widetext}
\begin{eqnarray}
    \begin{pmatrix}
        \chi_{ff}(\mathbf{0}) && \chi_{cf}(\mathbf{0}) 
        \\
        \chi_{cf}(\mathbf{0})  && \chi_{cc}(\mathbf{0}) 
    \end{pmatrix} &=&N_s \int_0^{\beta} {\rm d}\tau
    \left\langle \frac{N}{2}\begin{pmatrix}
        S_1(\tau)S_1(0) && S_1(\tau)s(0)
        \\
        S_1(\tau)s(0) && s(\tau)s(0)
    \end{pmatrix} \right\rangle_{N\rightarrow\infty},
    \nonumber\\
\chi_{ff}(\mathbf{0})  &=&
 \stackrel{\beta\rightarrow\infty}{=}
\frac{\rho J_K^2 \bar{V}_1^2}{2} 
\left( \frac{1}{\bar{\lambda}_1^2} - \frac{1}{J_K^2\bar{V}_1^2+ (\frac{-n_c}{2\rho}-\bar{\lambda}_1)^2}
\right),
\nonumber\\
\chi_{cc}(\mathbf{0})  &=&
 \stackrel{\beta\rightarrow\infty}{=}
\frac{\rho}{2} \frac{(-\frac{n_c}{2\rho}-\bar{\lambda}_1)^2}{(-\frac{n_c}{2\rho}-\bar{\lambda}_1)^2+J_K^2\bar{V}_1^2},
\nonumber\\
\chi_{cf}(\mathbf{0})  &=&
 \stackrel{\beta\rightarrow\infty}{=}
\frac{\rho}{2} \frac{J_K^2\bar{V}_1^2}{(-\frac{n_c}{2\rho}-\bar{\lambda}_1)^2+J_K^2\bar{V}_1^2},
\end{eqnarray}
\end{widetext}
where the susceptibilities have been computed at zero temperature ($\beta\rightarrow\infty$).
The $N\rightarrow\infty$ order parameters can be analytically computed
\begin{eqnarray}
    \bar{V}_1^2=\frac{n_c}{8\rho^2 J_K^2} \frac{1}{\cosh\left(\frac{1}{J_K\rho}\right)-1},\quad
    \bar{\lambda}_1 = \frac{n_c}{2\rho}\left(e^{\frac{1}{J_K\rho}}-1\right)^{-1}.
    \nonumber\\
\end{eqnarray}
It it is insightful to compare the strength of the RKKY-channel interaction with the interaction between $f_2$-moments that is generated by the residual ferromagnetic Hund-like exchange. From Eq.~\ref{fluc}, after integrating out the heavy-fermions, we obtain
\begin{eqnarray}
    \langle H_{\rm Hund} \rangle &=&-\frac{N}{2 N_S} \sum_{\mathbf{q}, i j}\mathcal{J}_{\rm Hund} (\mathbf{q})  e^{i \mathbf{q} \cdot(\mathbf{r}_i-\mathbf{r}_j)} \bar{S}_{2i}\bar{S}_{2j},
    \nonumber\\
    \mathcal{J}_{\rm Hund}(\mathbf{0}) &=&2\left(\frac{\bar{\lambda}_1}{J_K\bar{V}_1}\right)^4
    \frac{1}{\left(N\rho \log \frac{\bar{\lambda}_1}{T} \right)^2 }  \chi_{ff}(\mathbf{0}),
\end{eqnarray}
As $T\rightarrow 0$, $\mathcal{J}_{\rm Hund}(\mathbf{0})\rightarrow 0$ and so the RKKY interaction always dominates in the groundstate. However, we might also be interested in their competition at the critical temperature $T_{\rm RKKY}$ when magnetic order sets in. For $T=T_{\rm RKKY}\sim  \mathcal{J}_{\rm RKKY}(\mathbf{0})\sim \frac{1}{N^2} $, the residual ferromagnetic Hund-like interaction scales as ${J}_{\rm Hund}(\mathbf{0})\sim \frac{1}{ \left(N \log N\right)^2}$, i.e. it becomes negligible by comparison in the large-$N$ limit. Of course, once $N$ is no longer large, the interactions become comparable around $T_{\rm RKKY}$.

\end{document}